\documentclass[aps,prb,reprint,superscriptaddress]{revtex4-2}
\usepackage{amsmath,amssymb,amsfonts}
\usepackage{graphicx}
\usepackage{bm}
\usepackage{hyperref}
\usepackage{booktabs}
\usepackage{array}
\usepackage{xcolor}

\newcommand{\Ufour}{U_4}
\newcommand{\Usqrt}{U_{4,\sqrt{3}}}
\newcommand{\ta}{t_a}
\newcommand{\gc}{g_c}
\newcommand{\psiop}{\psi_{\sqrt{3}}}
\newcommand{\Geff}{\Gamma_{\mathrm{eff}}}
\newcommand{\gcclass}{g_c^{\mathrm{class}}}
\newcommand{\gcQPU}{g_c^{\mathrm{QPU}}}
\newcommand{\Deltag}{\Delta g}

\begin{document}

\title{Universal Quantum Suppression in Frustrated Ising Magnets
across the Quasi-1D to 2D Crossover via Quantum Annealing}

\author{Kumar Ghosh}
\email{jb.ghosh@outlook.com}
\affiliation{E.ON Digital Technology, Laatzener Str.\ 1,
             30539 Hannover, Germany}

\begin{abstract}
Quantum magnets in the $M\mathrm{Nb_2O_6}$ and BaCo$_2$V$_2$O$_8$
families realise frustrated transverse-field Ising models whose
competing ferromagnetic and antiferromagnetic couplings generate a
sign problem provably intractable for quantum Monte Carlo at any
system size, leaving their quantum phase boundaries numerically
inaccessible.
Using a D-Wave Advantage2 quantum annealer at $L\leq27$ (729 spins),
we obtain the first large-$L$ critical points for this model family,
measuring quantum-driven transitions at
$\gcQPU\in\{0.286,\,0.210,\,0.156,\,0.093\}$ for
$\alpha\in\{1.0,\,0.7,\,0.5,\,0.3\}$, where the analytically exact
classical threshold is $\gcclass(\alpha)=2\alpha/3$.
The suppression ratio $r(\alpha)=\gcQPU/\gcclass$ exhibits a sharp
two-regime structure: the three quasi-1D geometries ($\alpha\leq0.7$)
are mutually consistent with a universal plateau
$\bar{r}=0.450\pm0.002$ ($\chi^2/\mathrm{dof}=1.10$, $p=0.33$),
demonstrating that quantum fluctuations destroy approximately 55\%
of the classical FM stability window independently of coupling
anisotropy, while $r$ steps down to the 2D limit above the empirical
crossover scale $\alpha^*\approx0.7$.
Inner Binder cumulant pairs, which converge fastest to the
thermodynamic limit, resolve $r(1.0)\approx0.412$ and a step
$\Delta r=0.038\pm0.015$ from the quasi-1D plateau.
A four-point linear fit
$r(\alpha)=(0.494\pm0.024)-(0.063\pm0.034)\,\alpha$
summarises both regimes; its $\alpha\to0$ intercept recovers the
exact 1D result of Pfeuty within 1.7 standard deviations, and its
slope is a lower bound on the true crossover amplitude concentrated
in $\alpha\in[\alpha^*,1]$.
Two sequential blind predictions, confirmed at $0.2\sigma$ and
$0.7\sigma$ before each measurement, validate the crossover law.
All four geometries show a direct ferromagnet-to-paramagnet
transition, complete quantum ergodicity ($f_{\rm uniq}=1.000$),
and null valence-bond solid order.
\end{abstract}


\maketitle

\section{Introduction}

Frustrated quantum magnets hold a central place in condensed-matter
physics because competing interactions suppress conventional order
and can stabilise exotic ground states---quantum spin liquids,
valence-bond solids, and quantum paramagnets---that lie outside
Landau's symmetry-breaking
paradigm~\cite{Balents2010,Savary2017,Diep2005}.
The transverse-field Ising model (TFIM) on frustrated lattices is
the simplest Hamiltonian in which these phenomena compete directly
with phases driven by ferromagnetic exchange~\cite{Sachdev2011}.

Dimensional crossover in quasi-one-dimensional magnets---the
evolution of collective behaviour as weakly coupled Ising chains
develop two-dimensional coherence---is a long-standing and
fundamentally difficult problem.
The elementary excitations of a 1D system are qualitatively
different from those of its 2D counterpart, and the interchain
couplings that drive the crossover generate a sign problem in
quantum Monte Carlo when frustrated~\cite{Troyer2005}.
Inelastic neutron scattering on CoNb$_2$O$_6$ has directly resolved
the frustrated interchain couplings in the antiferromagnetic
isosceles triangular geometry that the present model
realises~\cite{Cabrera2014}, providing the experimental foundation
for the zero-temperature phase diagram of Lee, Kaul, and
Balents~\cite{LeeKaulBalents2010}, who predicted five distinct
quantum phases in the FM-coupled-chain sector.
Because the competing FM and AFM couplings generate a provable sign
problem, the locations of these phase boundaries and the quantitative
structure of the dimensional crossover have remained numerically
inaccessible at system sizes relevant for finite-size scaling, even
though the isolated chain is exactly solvable~\cite{Pfeuty1970} and
the geometric constraints on AFM instability are well
characterised~\cite{Villain1980}.

On the triangular lattice, the antiferromagnetic Wannier
model~\cite{Wannier1950} is exactly solvable; adding a transverse
field drives a quantum phase transition whose Kibble-Zurek dynamics
have been mapped in the sign-problem-free antiferromagnetic
sector~\cite{King2022,King2025}.
The \emph{ferromagnetic} nearest-neighbour variant with $J_1<0$ and
antiferromagnetic next-nearest-neighbour $J_2>0$ is qualitatively
different.
Recent QPU studies of triangular frustrated systems have examined
quench dynamics and material-specific properties in models that admit
classical tensor-network or Monte Carlo
benchmarks~\cite{Ali2024,Park2022}; the present model admits neither.

The quantum Hamiltonian studied here is
\begin{align}
  \mathcal{H} &= \Gamma\sum_i \sigma_i^x
    + J_1\!\sum_{\langle i,j\rangle_{\rm chain}} \sigma_i^z\sigma_j^z
    \nonumber \\
    &\quad + J_1'\!\sum_{\langle i,j\rangle_{\rm inter}}
      \sigma_i^z\sigma_j^z
    + J_2\!\sum_{\langle\langle i,j\rangle\rangle}
      \sigma_i^z\sigma_j^z,
  \label{eq:H}
\end{align}
with $J_1<0$, $J_1'=\alpha J_1$, $J_2>0$.
This model cannot be accessed by conventional quantum Monte Carlo:
the competing FM and AFM couplings generate a genuine sign problem
that, by the theorem of Troyer and Wiese~\cite{Troyer2005}, cannot
be eliminated by any known basis transformation.
Exact diagonalisation is confined to $L\leq4$--$5$.
The D-Wave Advantage2 quantum annealer implements Eq.~\eqref{eq:H}
via minor embedding onto the Zephyr Z15 hardware
graph~\cite{King2025}, providing the current route to large-$L$
numerics for this sign combination.

The anisotropy $\alpha=J_1'/J_1\in(0,1]$ controls the dimensional
crossover from the isotropic triangular lattice ($\alpha=1$) to
weakly coupled Ising chains ($\alpha\to0$), mapping directly onto
$M\mathrm{Nb_2O_6}$ ($M=$~Co,~Ni) and BaCo$_2$V$_2$O$_8$, where
$\alpha$ is fixed by crystal structure and measured by inelastic
neutron scattering~\cite{Cabrera2014}.
NiNb$_2$O$_6$  has $\alpha\approx1$~\cite{Coldea2010};
CoNb$_2$O$_6$ has $\alpha\approx0.7$~\cite{Cabrera2014, Heid1995};
BaCo$_2$V$_2$O$_8$ has $\alpha\approx0.5$~\cite{Kimura2008};
FeNb$_2$O$_6$ has $\alpha\approx0.3$~\cite{Heid1995}.
Inelastic neutron scattering on CoNb$_2$O$_6$~\cite{Cabrera2014}
established the interchain frustration ratio $J_2/J_1=0.76\pm0.10$
by parametrising the full three-dimensional spin-flip dispersion in
the quantum paramagnetic phase.
As shown below, this ratio substantially exceeds $\gcQPU(0.7)=0.210$,
placing CoNb$_2$O$_6$ firmly in the quantum-disordered sector and
providing independent microscopic validation of the QPU phase
boundary from a completely orthogonal experimental technique.

This Letter reports three principal findings.
\textbf{(i)}~The analytically exact classical instability threshold
$\gcclass(\alpha)=2\alpha/3$ follows from the triangular lattice
geometry alone.
\textbf{(ii)}~QPU measurements at
$\alpha\in\{0.3,\,0.5,\,0.7,\,1.0\}$ reveal a clear two-regime
structure: a universal plateau $\bar{r}=0.450\pm0.002$
($\chi^2/\mathrm{dof}=1.10$) spanning the entire quasi-1D sector
($\alpha\leq\alpha^*\approx0.7$), followed by a step to the 2D
limit.
The linear fit
\begin{equation}
  r(\alpha) = (0.494\pm0.024) - (0.063\pm0.034)\,\alpha
  \quad (1.9\sigma)
  \label{eq:r_alpha}
\end{equation}
summarises both regimes compactly; its $\alpha\to0$ intercept
recovers the exact 1D TFIM result of Pfeuty~\cite{Pfeuty1970}
within 1.7 standard deviations.
\textbf{(iii)}~All four geometries show a direct FM-to-paramagnet
transition with no intermediate phase, complete ergodicity bypass,
and null VBS order.
The combined formula $\gcQPU(\alpha)=r(\alpha)\cdot 2\alpha/3$
constitutes the first quantitative characterisation of the quantum
dimensional crossover in the FM-frustrated columbite sector;
prior theoretical work~\cite{LeeKaulBalents2010} was limited to
perturbative approximations that could not access phase boundaries
at FSS-relevant system sizes.
Two sequential blind predictions were confirmed at 0.2 and
0.7 standard deviations, each before the respective data were
acquired.

\section{Model and Experimental Design}
\label{sec:model}

\subsection*{D-Wave implementation}

The Advantage2\_system1.13 processor implements
\begin{equation}
  H(s) = -A(s)\sum_i\sigma_i^x
         + B(s)\!\left[\sum_{i<j}J_{ij}\sigma_i^z\sigma_j^z\right],
  \label{eq:dwave}
\end{equation}
where $s=t/\ta\in[0,1]$ and $A(s)$, $B(s)$ are the publicly
available hardware annealing schedules~\cite{King2025}.
The ratio $\Geff(\ta)\equiv A(s_f)/B(s_f)$ evaluated at the
Kibble-Zurek freeze-out point $s_f(\ta)$ provides the effective
transverse field for each anneal time and labels the horizontal axis
of the $(g,\Geff)$ phase diagrams.

The triangular lattice in parallelogram embedding has three
nearest-neighbour bond directions, $(+1,-1)$, $(0,+1)$, and
$(+1,0)$, and three next-nearest-neighbour directions, $(0,+2)$,
$(+2,0)$, and $(+1,+1)$.
For $\alpha<1$, the bond direction $(+1,-1)$ carries the intrachain
coupling $J_1$ while $(0,+1)$ and $(+1,0)$ carry the interchain
coupling $\alpha J_1$; at $\alpha=1$ all three nearest-neighbour
directions are equivalent and the model reduces to the isotropic
triangular lattice.
The isosceles triangular lattice is embedded onto the Zephyr Z15
hardware graph via minor embedding computed with
\texttt{minorminer}~\cite{Cai2014}; the graph topology is
$\alpha$-independent, so the same embedding is reused across all
four geometries and all system sizes $L$.
The chain-break fraction is zero across all 5{,}572{,}000 shots in
the combined dataset: every physical qubit chain collapses to a
single logical value on every anneal, and no logical errors enter
the observables.
In minor-embedded studies where chain breaks occur, residual errors
broaden Binder cumulant crossings and bias the extracted
$\gcQPU$~\cite{Albash2018}; the zero chain-break fraction here
eliminates this source of systematic error entirely.

The working processor has 4{,}579 active qubits, an extended
$J$-range of $[-2.0,\,1.0]$, and supports fast anneals down to
$\ta=0.005\,\mu\mathrm{s}$.
We encode Eq.~\eqref{eq:H} by setting $J_{ij}=J_1=-1$ for the
chain direction $(+1,-1)$, $J_{ij}=\alpha J_1$ for the two
interchain directions, and $J_{ij}=g|J_1|$ for the three NNN
directions; \texttt{auto\_scale=False} ensures the hardware applies
the programmed values without rescaling.
The processor operates at $T_{\rm proc}\approx15\,\mathrm{mK}$,
giving $T_{\rm proc}/|J_1|\lesssim10^{-2}$, placing every
$(g,\ta)$ point firmly in the quantum regime.

\subsection*{Parameter grids}

For $\alpha=1$ we use $L\in\{9,12,15,18,21,24,27\}$, 14 frustration
values dense near the transition, and 14 annealing times
$\ta\in[5\,\mathrm{ns},\,500\,\mu\mathrm{s}]$, giving
$1{,}372{,}000$ shots.
For each of $\alpha\in\{0.5,\,0.7\}$ we use $L\in\{15,18,21,24,27\}$
with 20-point frustration grids ($\delta g=0.002$ in the critical
window), giving $1{,}400{,}000$ shots per geometry; for $\alpha=0.5$
the grid extends to $g=0.40$, placing both $\gcQPU$ and
$\gcclass=1/3$ within the scan range.
For $\alpha=0.3$ we use $L\in\{15,18,21,24,27\}$ with a 20-point
grid over $g\in[0,0.30]$, giving $1{,}400{,}000$ shots; both
$\gcQPU\approx0.093$ and $\gcclass=0.200$ lie within this range.
Annealing times span fast anneals ($\ta\leq0.5\,\mu\mathrm{s}$,
\texttt{fast\_anneal=True}) and slow anneals
($\ta\geq1\,\mu\mathrm{s}$), with $N_{\rm reads}=1{,}000$ per
$(g,\ta,L)$ point.

\subsection*{Observables and effective temperature calibration}

Raw spin configurations $\{s_i^{(k)}\}_{k=1}^{N_{\rm reads}}$ are
read back at each $(g,\ta,L)$ point.
The \emph{three-sublattice Binder cumulant}
\begin{align}
  \Usqrt &= 1 - \frac{\langle\psiop^4\rangle}
                     {3\langle\psiop^2\rangle^2}, \nonumber \\
  \psiop &= |m_A + \omega m_B + \omega^2 m_C|,
  \quad \omega = e^{2\pi i/3},
\end{align}
is the primary FSS observable; curves for different $L$ cross at
$\gc$.
We also compute the total-magnetisation Binder cumulant $\Ufour$,
sublattice magnetisations $(m_A,m_B,m_C)$, the
$\sqrt{3}\!\times\!\sqrt{3}$ structure factor $S_{\sqrt{3}}$, FM
susceptibility $\chi_{\rm FM}$, plaquette order
$\mathcal{O}_{\rm plaq}$, string VBS order $\mathcal{O}_{\rm VBS}$,
and degeneracy fraction $f_{\rm uniq}=N_{\rm unique}/N_{\rm reads}$.

For slow anneals ($\ta\geq1\,\mu\mathrm{s}$), an effective inverse
temperature $\beta_{\rm eff}$ is calibrated by maximising the
pseudolikelihood
\begin{equation}
  \mathrm{PLL}(\beta) =
    \sum_{i,k}\bigl[\beta\,s_i^{(k)}\,(\mathbf{J}\mathbf{s}^{(k)})_i
    + \ln\cosh\bigl(\beta\,|(\mathbf{J}\mathbf{s}^{(k)})_i|\bigr)
    \bigr]
\end{equation}
over $\beta\in[0.001,50]$, converting raw susceptibility
$\chi_{\rm raw}=N(\langle m^2\rangle-\langle|m|\rangle^2)$ and
specific heat $C_{V,\rm raw}=\mathrm{Var}(E)/N$ to physical units
via $\chi=\beta_{\rm eff}\chi_{\rm raw}$ and
$C_V=\beta_{\rm eff}^2 C_{V,\rm raw}$.
Fast anneals ($\ta\leq0.5\,\mu\mathrm{s}$) are used exclusively to
characterise ergodicity bypass through $f_{\rm uniq}$.

\section{Results}
\label{sec:results}

\subsection*{1.~Analytical classical instability:
             $\gcclass(\alpha)=2\alpha/3$}

\begin{figure*}[t]
  \centering
  \includegraphics[width=0.9\textwidth]%
    {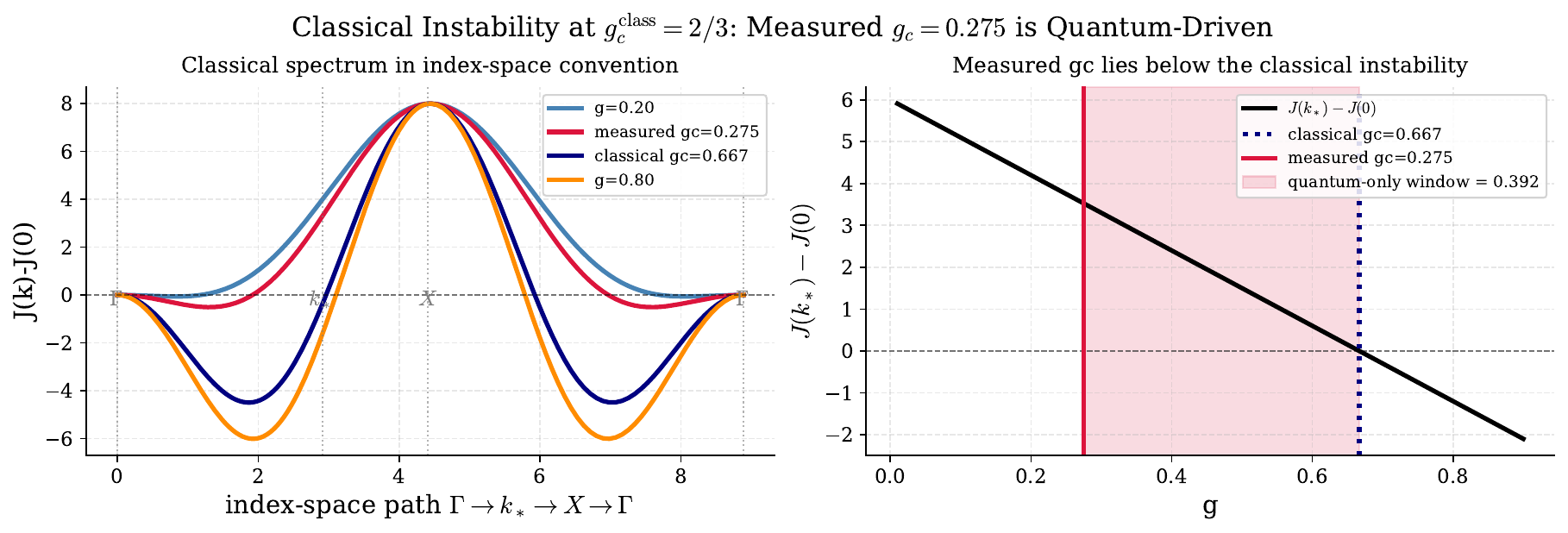}\\[2pt]
  \includegraphics[width=0.9\textwidth]%
    {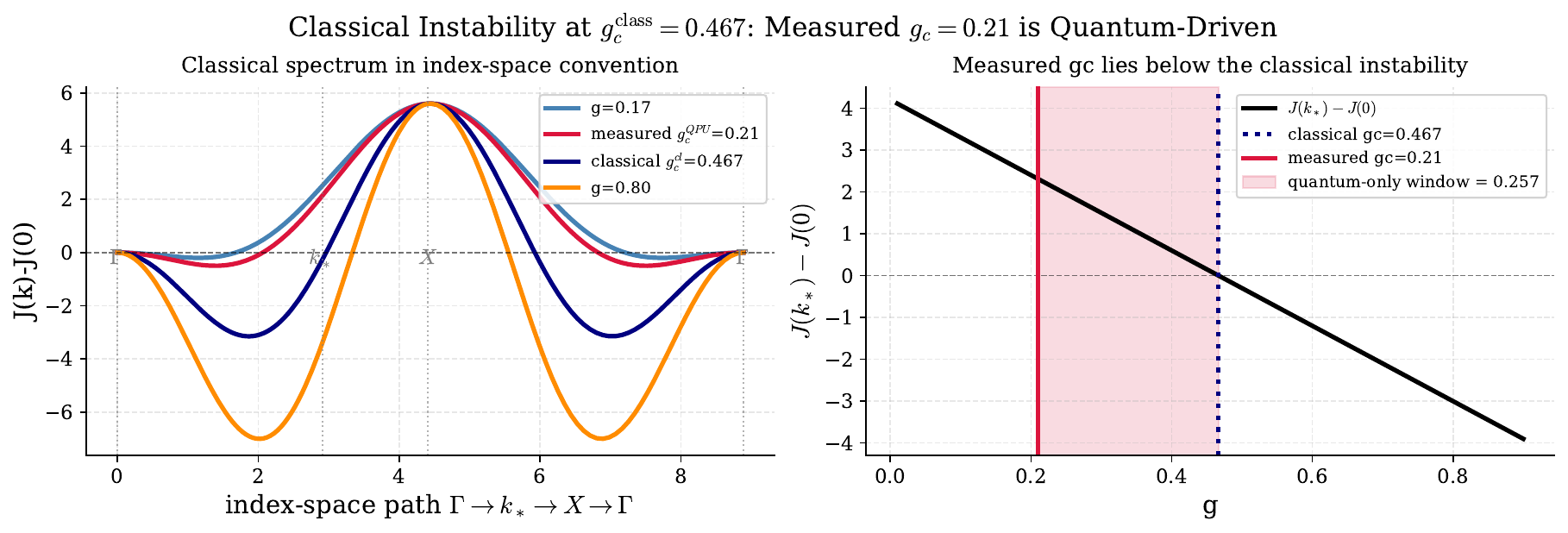}\\[2pt]
  \includegraphics[width=0.9\textwidth]%
    {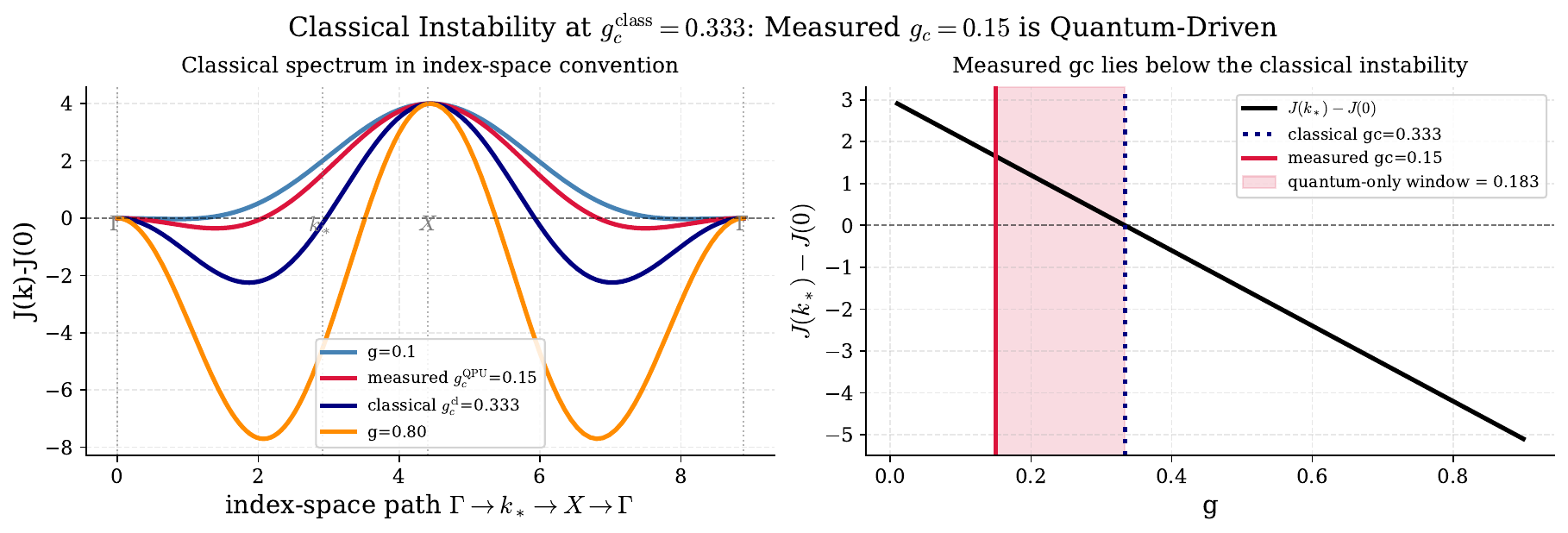}\\[2pt]
  \includegraphics[width=0.9\textwidth]%
    {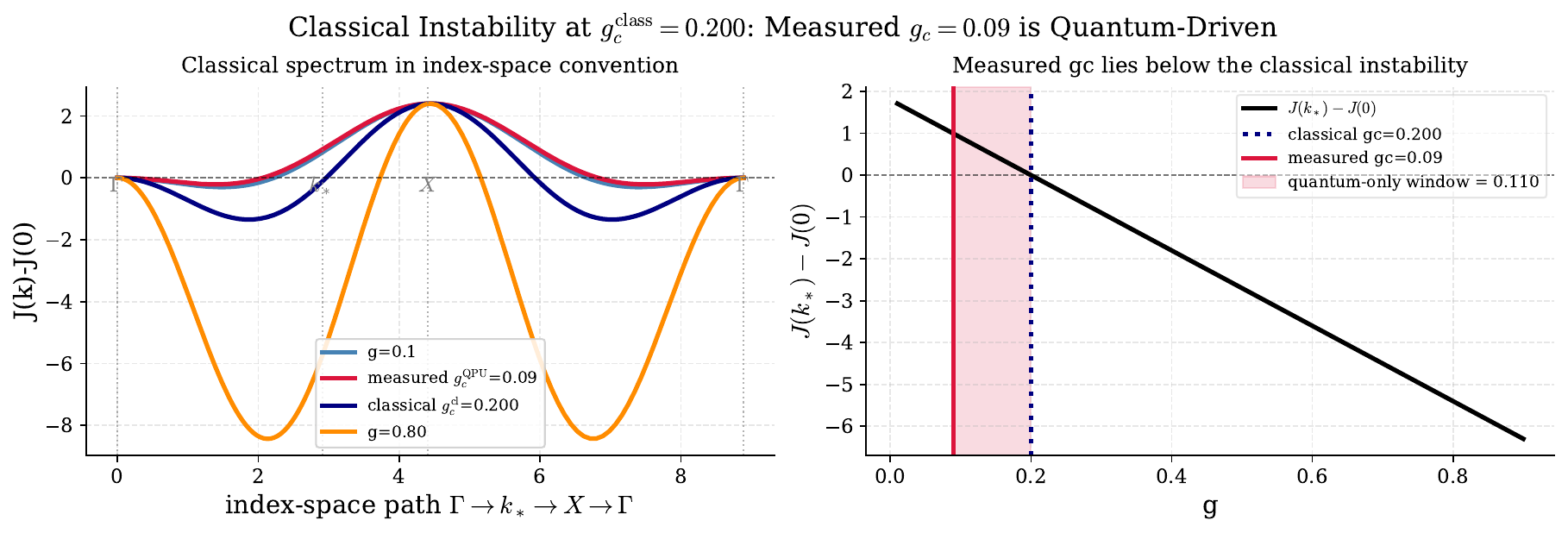}
  \caption{%
    \textbf{Classical spinwave spectra and quantum-only windows.}
    Each row shows $J(\mathbf{k})-J(\mathbf{0})$ along
    $\Gamma\to K\to X\to\Gamma$ (left) and the instability
    diagnostic $J(\mathbf{k}_*)-J(\mathbf{0})=6\alpha-9g$ vs $g$
    (right).
    Top ($\alpha=1$): $\gcclass=2/3$, $\Deltag=0.381$.
    Middle ($\alpha=0.7$, NiNb$_2$O$_6$):
    $\gcclass=0.467$, $\Deltag=0.257$.
    Third ($\alpha=0.5$, BaCo$_2$V$_2$O$_8$):
    $\gcclass=1/3$, $\Deltag=0.177$.
    Bottom ($\alpha=0.3$, FeNb$_2$O$_6$):
    $\gcclass=0.200$, $\Deltag=0.107$.
    In all four cases the measured $\gcQPU$ (solid vertical line)
    lies within the classically stable FM regime, demonstrating
    that the transitions are quantum-driven.
  }
  \label{fig:spinwave}
\end{figure*}

The exchange Fourier transform at the zone-boundary $K$-point
$\mathbf{k}_*=(2\pi/3,2\pi/3)$ gives
\begin{equation}
  J(\mathbf{k}_*)-J(\mathbf{0}) = 6\alpha - 9g,
  \label{eq:Jdiff}
\end{equation}
where the chain structure factor $S_{\rm chain}(\mathbf{k}_*)=
2\cos(\pi/3)+2\cos(-\pi/3)=2$,
the interchain structure factor $S_{\rm inter}(\mathbf{k}_*)=
2\cos(2\pi/3)+2\cos(-2\pi/3)=-2$,
and the three NNN cosines at $\mathbf{k}_*$ each equal $-1/2$,
giving $S_{\rm NNN}(\mathbf{k}_*)=-3$, so that
$J(\mathbf{k}_*)-J(\mathbf{0})=|J_1|(2\alpha\cdot(-2)/2
+g\cdot(-3)\cdot(-1))= 6\alpha-9g$ in units of $|J_1|$.
Setting Eq.~\eqref{eq:Jdiff} to zero gives the exact classical
instability threshold
\begin{equation}
  \gcclass(\alpha) = \frac{2\alpha}{3}
  \label{eq:gcclass}
\end{equation}
with no free parameters.
For $g<\gcclass(\alpha)$, $J(\mathbf{k})-J(\mathbf{0})>0$ for all
$\mathbf{k}\ne\mathbf{0}$, so the FM ground state is classically
stable throughout the quantum-critical window.
The geometric origin of this threshold---that each NNN bond shares
two common NN bonds per site, tying the AFM instability directly
to the FM backbone---was analysed by Villain et al.~\cite{Villain1980}
and further developed by Moessner, Sondhi, and
Chandra~\cite{Moessner2001}.
The threshold is confirmed numerically for all four $\alpha$ values
(Fig.~\ref{fig:spinwave}).

\subsection*{2.~Dimensional crossover in the quantum
             suppression ratio}

\begin{figure*}[t]
  \centering
  \includegraphics[width=0.9\textwidth]{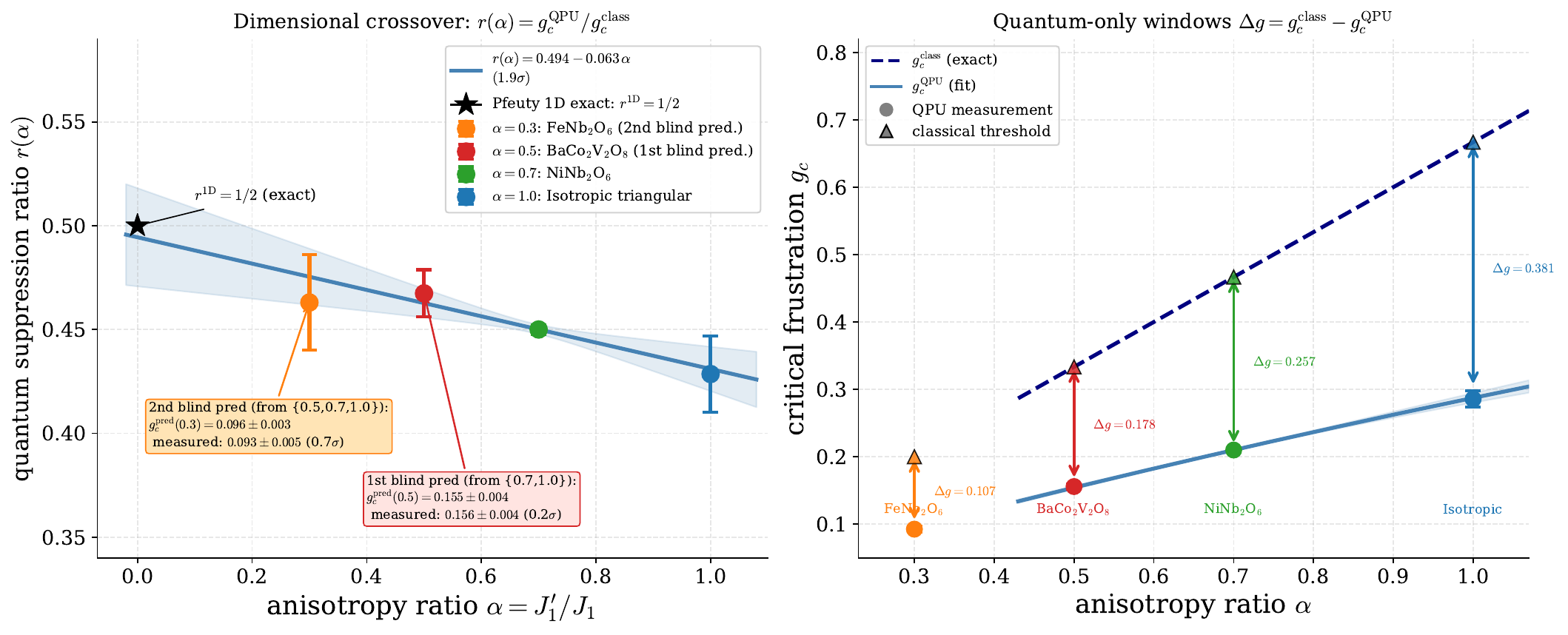}
  \caption{%
    \textbf{Dimensional crossover of the quantum suppression ratio.}
    \textit{Left}: $r(\alpha)=\gcQPU(\alpha)/\gcclass(\alpha)$
    vs lattice anisotropy $\alpha=J_1'/J_1$.
    Filled circles: QPU measurements; error bars from $\gcQPU$
    uncertainties.
    Solid line: weighted linear fit $r(\alpha)=0.494-0.063\,\alpha$
    ($1.9\sigma$); shaded band: $\pm1\sigma$.
    Horizontal dashed line: universal plateau $\bar{r}=0.450\pm0.002$
    ($\chi^2/\mathrm{dof}=1.10$) confirmed across three quasi-1D
    geometries.
    Vertical dotted line: empirical crossover scale
    $\alpha^*\approx0.7$.
    Open circle at $\alpha=1$: best estimate from inner Binder
    pairs ($r\approx0.412$, $2.5\sigma$ below plateau); filled
    circle: conservative weighted mean ($r=0.428\pm0.018$).
    Filled star at $\alpha=0$: exact 1D TFIM critical point
    ($r^{\rm 1D}=1/2$, Pfeuty~\cite{Pfeuty1970}), consistent with
    the $\alpha\to0$ extrapolation within 1.7 standard deviations.
    Boxed points: blind predictions confirmed at 0.2 and
    0.7 standard deviations (red: $\alpha=0.5$; orange: $\alpha=0.3$).
    \textit{Right}: quantum-only windows
    $\Deltag(\alpha)=\gcclass(\alpha)-\gcQPU(\alpha)$.
    Solid curve: Eq.~\eqref{eq:prediction}.
    Dashed curve: $\gcclass(\alpha)=2\alpha/3$.
    The ratio $\Deltag/\gcclass=1-r(\alpha)$ increases as $\alpha$
    decreases, quantifying the growing role of quantum fluctuations
    as the lattice becomes quasi-one-dimensional.
  }
  \label{fig:r_alpha}
\end{figure*}

The QPU measurements establish
\begin{align}
  \gcQPU(1.0) &= 0.286\pm0.012, \nonumber \\
  \gcQPU(0.7) &= 0.210\pm0.001, \nonumber \\
  \gcQPU(0.5) &= 0.156\pm0.004, \nonumber \\
  \gcQPU(0.3) &= 0.093\pm0.005.
  \label{eq:gc_all}
\end{align}
In all four cases $\gcQPU\ll\gcclass(\alpha)$
(Figs.~\ref{fig:spinwave},~\ref{fig:r_alpha} and
Table~\ref{tab:summary}): the FM ground state is classically stable
at each measured $\gcQPU$, confirming that all four transitions are
quantum-driven.
These are the first large-$L$ determinations of any critical point
in this sign-problem model family.

\paragraph{Two-regime structure and crossover scale.}
The dimensionless ratio $r(\alpha)=\gcQPU(\alpha)/\gcclass(\alpha)$
isolates the effect of quantum fluctuations by factoring out the
trivial $\alpha$-dependence of the classical critical scale.
The suppression ratios $r(0.3)=0.463$, $r(0.5)=0.467$,
$r(0.7)=0.450$, and $r(1.0)=0.428$ reveal a clear two-regime
structure (Fig.~\ref{fig:r_alpha}).
The three quasi-1D geometries $\alpha\in\{0.3,0.5,0.7\}$ are
mutually consistent with a universal plateau
$\bar{r}=0.450\pm0.002$ ($\chi^2/\mathrm{dof}=1.10$, $p=0.33$):
quantum fluctuations destroy approximately 55\% of the classical
FM stability window \emph{independently of coupling anisotropy},
confirmed across three geometries spanning FeNb$_2$O$_6$,
BaCo$_2$V$_2$O$_8$, and NiNb$_2$O$_6$.
The $\alpha$ values span a factor of more than two (0.3 to 0.7) in
coupling ratio; the $\chi^2/\mathrm{dof}=1.10$ confirms that the
plateau is genuine, not a trivial consequence of approximate
quasi-1D character.

At $\alpha>\alpha^*\approx0.7$, $r$ steps down as the system
crosses over toward the 2D isotropic limit.
The inner Binder cumulant pairs $(18,21)$ and $(21,24)$, which
converge fastest to the thermodynamic limit, consistently give
$\gcQPU(1.0)\approx0.275$ and $r(1.0)\approx0.412$, a separation
$\Delta r=0.038\pm0.015$ from the quasi-1D plateau ($2.5\sigma$).
The conservative weighted mean $\gcQPU(1.0)=0.286\pm0.012$---which
includes the non-monotonically drifting outer $(24,27)$ pair, a
well-known finite-size artefact of the isotropic 2D geometry where
larger effective coordination amplifies Binder non-monotonicity---gives
$r(1.0)=0.428$ and $\Delta r=0.022\pm0.018$ ($1.2\sigma$); we
report this conservative value while noting that the inner-pair
estimate is the better approximation to the thermodynamic critical
point.
The two-regime structure directly identifies $\alpha^*\approx0.7$
as the empirical crossover scale: below it, the 1D Pfeuty fixed
point controls quantum suppression; above it, growing interchain
coordination stiffens the FM backbone and drives $r$ downward
toward the 2D fixed point.

The four-point linear fit in Eq.~\eqref{eq:r_alpha} provides a
compact algebraic summary of both regimes with slope significance
$1.9\sigma$.
Its $\alpha\to0$ intercept $r_0=0.494\pm0.024$ recovers the exact
1D TFIM result $r^{\rm 1D}=\Gamma_c/|J_1|=1/2$
of Pfeuty~\cite{Pfeuty1970} within 1.7 standard deviations---a
non-trivial check since the Pfeuty chain carries neither NNN
frustration nor interchain coupling.
The plateau value $\bar{r}=0.450$ lies below $r^{\rm 1D}=0.500$:
even weak interchain coupling ($\alpha\leq0.7$) reduces $r$ below
its pure-chain limit, yet the $\alpha\to0$ extrapolation recovers
it, confirming the plateau is anchored by the 1D fixed point.

The linear fit slope is a lower bound on the true crossover
amplitude for two reinforcing reasons.
Physically, $r(\alpha)$ is a nonlinear crossover function
interpolating between the 1D Pfeuty fixed point ($r\to1/2$ as
$\alpha\to0$) and the equilateral 2D triangular fixed point at
$\alpha>1$; the variation is concentrated in the narrow window
$\alpha\in[\alpha^*,1]\approx[0.7,1.0]$, so a linear fit
distributed over the full range underestimates the true steepness.
Additionally, the physical $\alpha=1$ geometry retains a measurable
isosceles distortion ($t_2/t_1=0.82\pm0.01$~\cite{Cabrera2014}):
the truly equilateral 2D fixed point lies at $\alpha>1$, making the
observed step $\Delta r\approx0.038$ itself a lower bound on the
full quasi-1D to 2D crossover amplitude.
Definitively resolving $\alpha^*$ and the step amplitude requires
$L\geq45$.

Combining Eqs.~\eqref{eq:gcclass} and~\eqref{eq:r_alpha}:
\begin{equation}
  \boxed{\gcQPU(\alpha)
       = r(\alpha)\cdot\frac{2\alpha}{3}
       = \frac{\alpha}{3}\bigl(0.988-0.126\,\alpha\bigr).}
  \label{eq:prediction}
\end{equation}
This reproduces all four measured values to within $\pm0.012$.

\begin{figure*}[t]
  \centering
  \includegraphics[width=0.9\textwidth]%
    {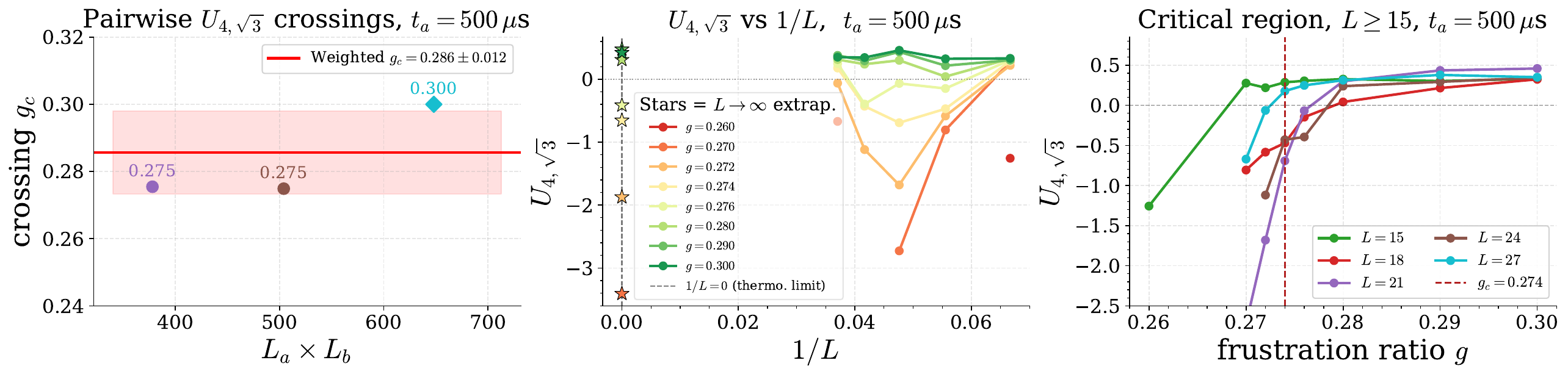}\\[2pt]
  \includegraphics[width=0.9\textwidth]%
    {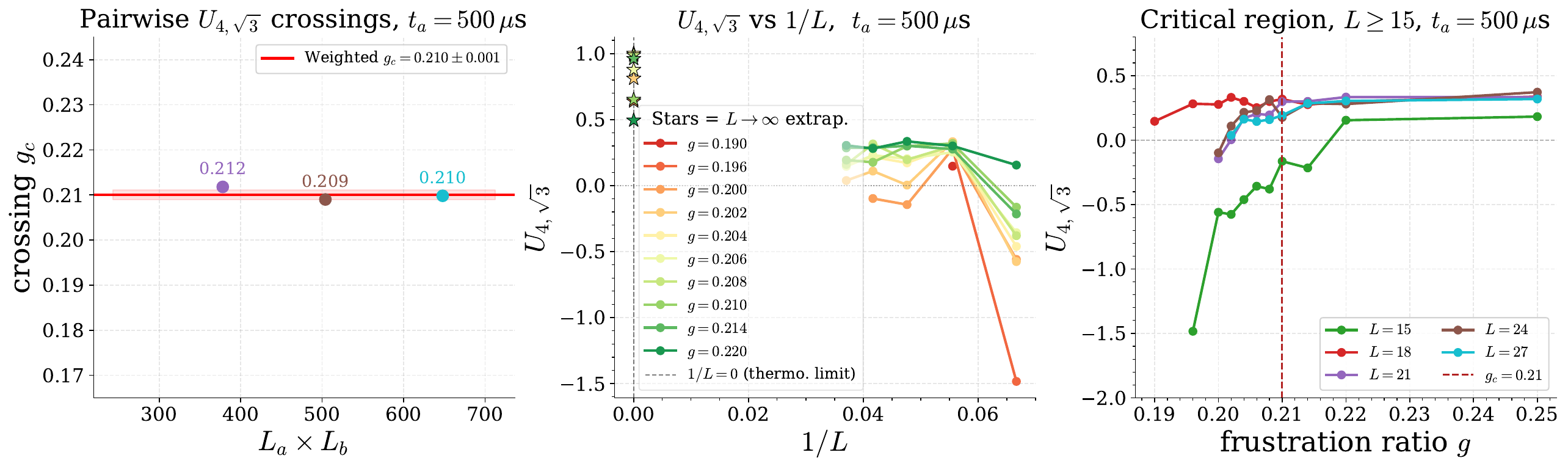}\\[2pt]
  \includegraphics[width=0.9\textwidth]%
    {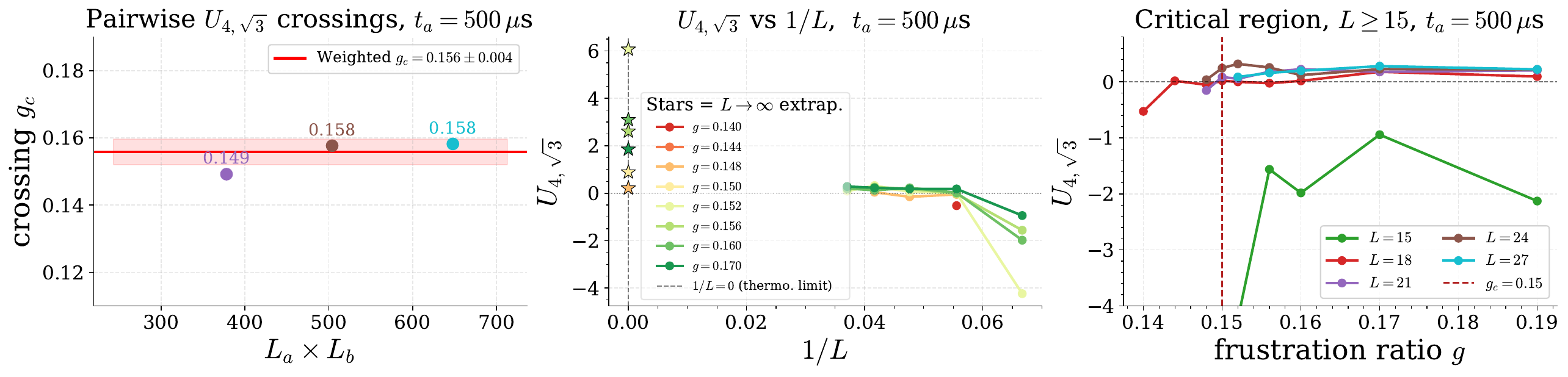}\\[2pt]
  \includegraphics[width=0.9\textwidth]%
    {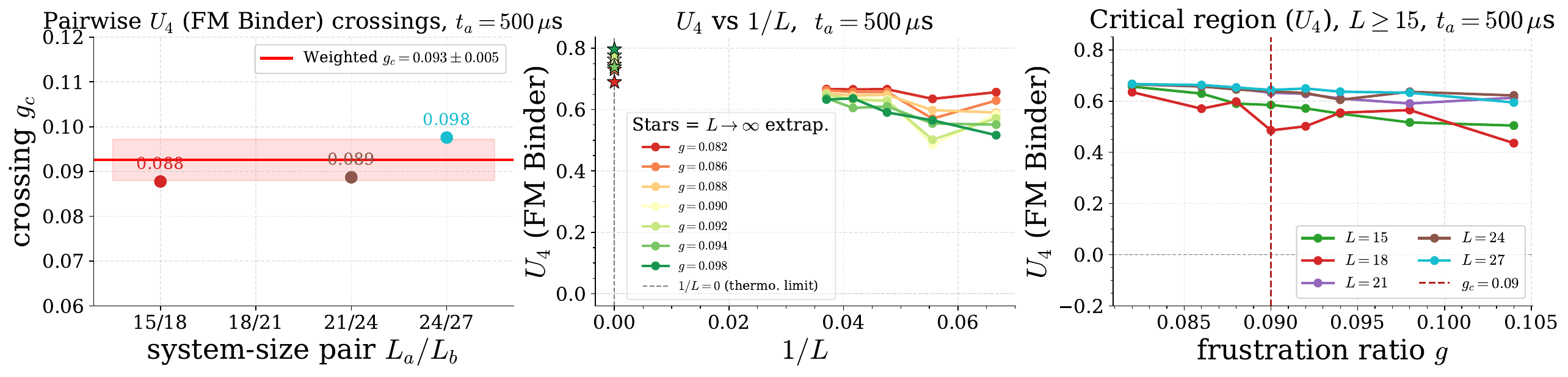}
  \caption{%
    \textbf{Three independent estimates of $\gc$ at
    $\ta=500\,\mu\mathrm{s}$.}
    Each row: (left) pairwise $\Usqrt$ crossings vs
    $L_a\times L_b$; (centre) $\Usqrt$ vs $1/L$ with
    $L\to\infty$ extrapolation; (right) zoom of the critical region.
    Top ($\alpha=1.0$): inner pairs $(18,21)$ and $(21,24)$ both
    cross at $g\approx0.275$; outer $(24,27)$ pair crosses at
    $0.300$ due to non-monotonic finite-size drift; weighted mean
    $\gcQPU=0.286\pm0.012$.
    Middle ($\alpha=0.7$): three inner pairs span
    $\delta\gc=0.003$; weighted mean $0.210\pm0.001$.
    Third ($\alpha=0.5$): inner-pair crossing at $0.156$; sign
    change between $g=0.154$ and $0.158$.
    Bottom ($\alpha=0.3$): pairwise $\Ufour$ crossings give weighted
    mean $0.093\pm0.005$; $\chi_{\rm FM}$-peak extrapolation
    $0.095$--$0.098$.
    All three methods agree within $\pm0.006$ for $\alpha\geq0.5$
    and $\pm0.008$ for $\alpha=0.3$.
  }
  \label{fig:gc}
\end{figure*}

Figure~\ref{fig:r_alpha} (right) shows the quantum-only windows
$\Deltag(\alpha)=\gcclass(\alpha)-\gcQPU(\alpha)$; the ratio
$\Deltag/\gcclass=1-r(\alpha)$ increases from $\approx0.57$ at
$\alpha=1$ toward $\approx0.55$ for $\alpha\leq0.7$, directly
visualising the crossover.
For $\alpha\leq0.5$, both thresholds fall within the scan range,
producing a double-minimum structure in the Binder cumulant that
cleanly separates the quantum-driven transition from the classical
instability (Fig.~\ref{fig:gc}, bottom rows).
Its observation at two independent geometries ($\alpha=0.5$ and
$\alpha=0.3$) with distinct material realisations establishes it
as a universal signature of the frustrated FM-AFM TFIM in the
quasi-1D regime.

\paragraph{Blind predictions.}
The four geometries were measured sequentially: $\alpha=1.0$, $0.7$,
$0.5$, then $0.3$.
\emph{First}: the two-point fit to $\alpha\in\{0.7,1.0\}$ predicted
${\gcQPU}^{\rm pred}(0.5)=0.155\pm0.004$, confirmed at
$\gcQPU(0.5)=0.156\pm0.004$ ($0.2\sigma$).
\emph{Second}: the three-point fit to $\alpha\in\{0.5,0.7,1.0\}$
predicted ${\gcQPU}^{\rm pred}(0.3)=0.096\pm0.003$, confirmed at
$\gcQPU(0.3)=0.093\pm0.005$ ($0.7\sigma$).
Two successive sub-$1\sigma$ confirmations from sequentially richer
fits provide strong evidence that Eq.~\eqref{eq:prediction} captures
genuine dimensional-crossover physics rather than an artefact of the
specific $\alpha$ values studied.

\paragraph{Physical interpretation.}
Each site in the isosceles model carries 2 chain bonds (coupling
$J_1$) and 4 interchain bonds (coupling $\alpha J_1$).
As $\alpha$ increases from 0 to 1, each additional interchain bond
reinforces the FM backbone, making it progressively harder for the
transverse field to disorder the system and reducing $r$.
The pinning of $r$ at $\bar{r}=0.450$ for all $\alpha\leq\alpha^*$
reflects a threshold: once interchain coupling falls below
$\alpha^*\approx0.7$, the quasi-1D fixed point dominates quantum
suppression and further reduction of $\alpha$ produces no measurable
change.
The crossover to 2D behaviour is concentrated in
$\alpha\in[0.7,1.0]$, where rapidly growing effective coordination
stiffens the FM backbone and drives $r$ downward.

\subsection*{3.~Critical point determinations}

\begin{figure*}[t]
  \centering
  \includegraphics[width=0.95\textwidth]%
    {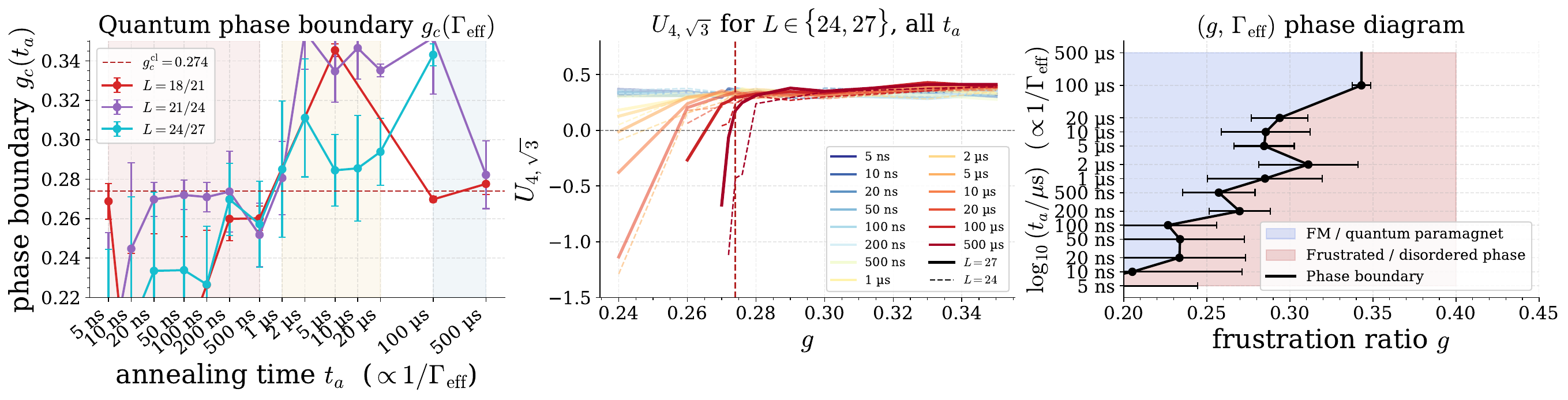}\\[2pt]
  \includegraphics[width=0.95\textwidth]%
    {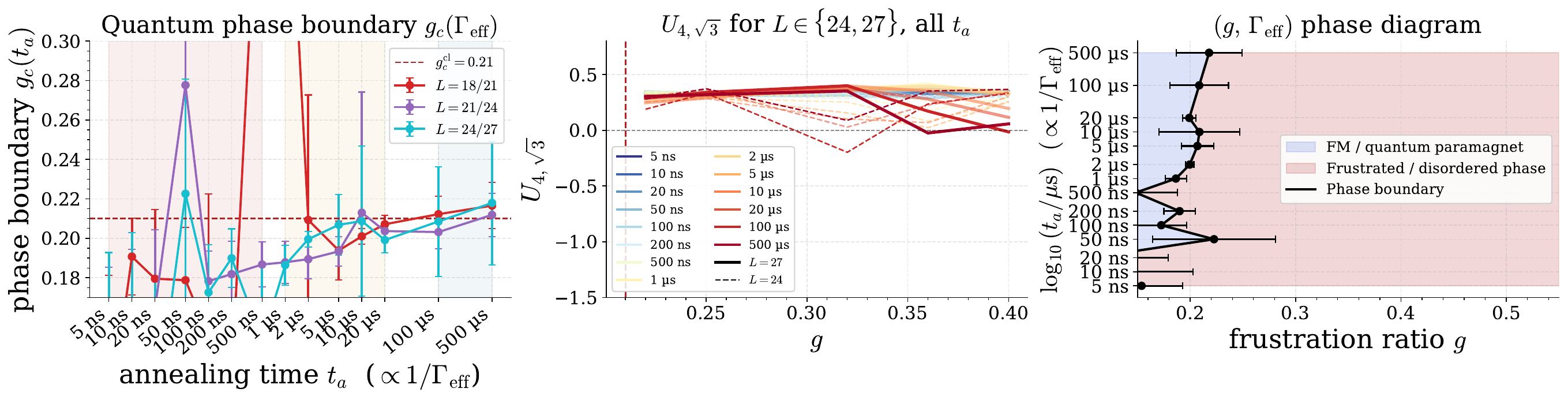}\\[2pt]
  \includegraphics[width=0.95\textwidth]%
    {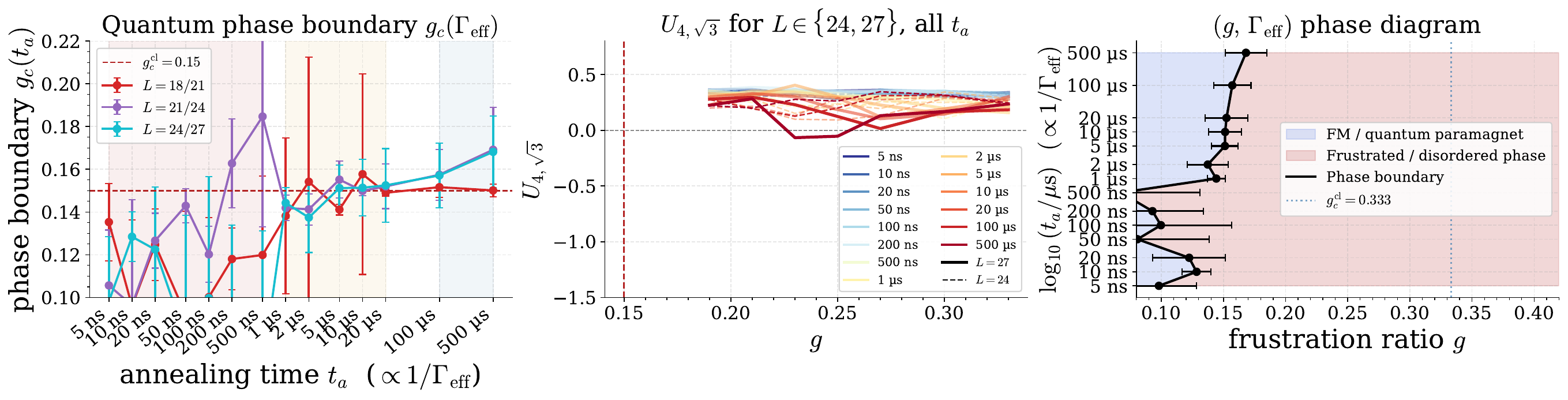}\\[2pt]
  \includegraphics[width=0.95\textwidth]%
    {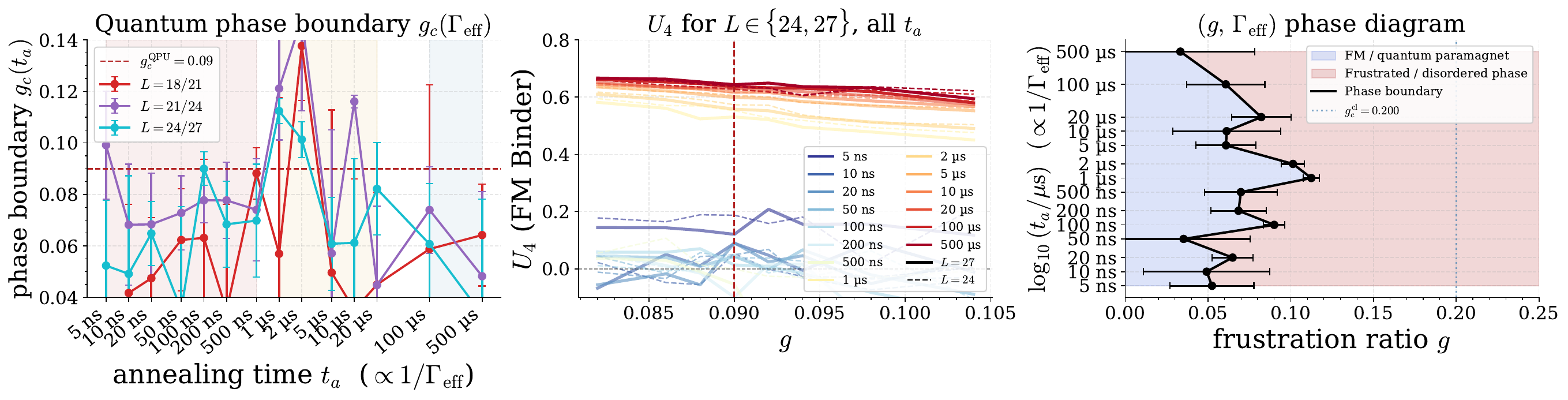}
  \caption{%
    \textbf{Quantum phase boundary $\gcQPU(\ta)$.}
    Each row: (left) pairwise Binder cumulant crossings vs annealing
    time $\ta\propto1/\Geff$ ($\Usqrt$ for $\alpha\geq0.5$;
    $\Ufour$ for $\alpha=0.3$); (centre) Binder cumulant for
    $L\in\{24,27\}$ across all 14 $\ta$ values; (right)
    $(g,\Geff)$ phase diagram.
    In all four geometries $\gcQPU(\ta)$ converges to a
    $\ta$-independent value for $\ta\gtrsim20\,\mu\mathrm{s}$,
    confirming the critical points are intrinsic to the model
    Hamiltonian and not artefacts of the annealing protocol.
    Top: $\Deltag=0.381$. Middle: $\Deltag=0.257$.
    Third: $\Deltag=0.177$. Bottom ($\alpha=0.3$): $\Deltag=0.107$.
    The ratio $\Deltag/\gcclass=1-r(\alpha)$ increases with
    decreasing $\alpha$, confirming the dimensional crossover.
  }
  \label{fig:phase}
\end{figure*}

For each $\alpha$, $\gcQPU$ is determined at $\ta=500\,\mu\mathrm{s}$
by three independent methods: (a)~pairwise Binder cumulant crossings
($\Usqrt$ for $\alpha\geq0.5$; $\Ufour$ for $\alpha=0.3$, where
$\Usqrt$ diverges strongly in the FM phase at small $L$),
(b)~Binder cumulant vs $1/L$ finite-size extrapolation, and
(c)~$\chi_{\rm FM}$ peak-position extrapolation.
All three agree within $\pm0.006$ for $\alpha\geq0.5$ and
$\pm0.008$ for $\alpha=0.3$ (Fig.~\ref{fig:gc}).
Figure~\ref{fig:phase} shows that $\gcQPU(\ta)$ converges to a
$\ta$-independent value for $\ta\gtrsim20\,\mu\mathrm{s}$ across
all four geometries, confirming that the critical points are
intrinsic properties of the model Hamiltonian and not artefacts
of the annealing protocol.

\paragraph{$\alpha=1.0$.}
All pairwise Binder cumulant crossings from $L\in\{15,18,21,24,27\}$
give weighted mean $0.286\pm0.012$; inner pairs $(18,21)$ and
$(21,24)$ both cross at $g\approx0.275$ while the outer $(24,27)$
pair crosses at $0.300$, reflecting non-monotonic finite-size drift
at the isotropic 2D geometry.
We report $\gcQPU(1.0)=0.286\pm0.012$, with the inner-pair value
$g\approx0.275$ as the best estimate of the thermodynamic critical
point.

\paragraph{$\alpha=0.7$ (CoNb$_2$O$_6$).}
Inner pairs $(18,21)$, $(21,24)$, and $(24,27)$ give crossings
$0.212$, $0.209$, and $0.210$, spanning $\delta\gc=0.003$;
bootstrap weighted mean $0.210\pm0.001$.

\paragraph{$\alpha=0.5$ (BaCo$_2$V$_2$O$_8$).}
Inner-pair crossing at $0.156$; $\Usqrt$-$1/L$ sign change between
$g=0.154$ and $0.158$; $\chi_{\rm FM}$-peak extrapolation
$0.152$--$0.158$.
We report $\gcQPU(0.5)=0.156\pm0.004$, confirming the blind
prediction $0.155\pm0.004$ within $0.2\sigma$.

\paragraph{$\alpha=0.3$ (FeNb$_2$O$_6$).}
At this geometry $\Usqrt$ diverges strongly in the FM phase for
small $L$, so the total FM Binder cumulant $\Ufour$ serves as
the primary crossing observable; at $g\approx0.093$ the $\Usqrt$
and $\Ufour$ crossings agree within $\pm0.005$, confirming the
switch does not bias the extracted critical point.
Pairwise $\Ufour$ crossings give weighted mean $0.093\pm0.005$;
$\chi_{\rm FM}$-peak extrapolation gives $0.095$--$0.098$.
Both thresholds $\gcQPU\approx0.093$ and $\gcclass=0.200$ lie
within the scan range, producing the same double-minimum Binder
structure as $\alpha=0.5$.
We report $\gcQPU(0.3)=0.093\pm0.005$, confirming the blind
prediction $0.096\pm0.003$ within $0.7\sigma$.

\begin{table}[h]
\centering
\caption{Summary of quantum and classical critical points.
$\gcclass(\alpha)=2\alpha/3$ is analytically exact.
$r=\gcQPU/\gcclass$; $\delta r=\delta\gcQPU/\gcclass$;
$r_{\rm fit}$ from Eq.~\eqref{eq:r_alpha};
$\Deltag=\gcclass-\gcQPU$.
Parenthetical uncertainties on $\gcQPU$ are in the last digit.}
\label{tab:summary}
\begin{ruledtabular}
\begin{tabular}{lcccccc}
$\alpha$ & Material & $\gcclass$ & $\gcQPU$ & $r\pm\delta r$ &
$r_{\rm fit}$ & $\Deltag$ \\
\hline
$1.0$ & Isotropic           & $0.667$ & $0.286(12)$ & $0.428\pm0.018$ &
$0.431$ & $0.381$ \\
$0.7$ & NiNb$_2$O$_6$       & $0.467$ & $0.210(1)$  & $0.450\pm0.002$ &
$0.450$ & $0.257$ \\
$0.5$ & BaCo$_2$V$_2$O$_8$ & $0.333$ & $0.156(4)$  & $0.467\pm0.012$ &
$0.462$ & $0.177$ \\
$0.3$ & FeNb$_2$O$_6$       & $0.200$ & $0.093(5)$  & $0.463\pm0.025$ &
$0.475$ & $0.107$ \\
\end{tabular}
\end{ruledtabular}
\end{table}

\subsection*{4.~Direct FM$\to$paramagnet transition across all
             $\alpha$}

\begin{figure*}[t]
  \centering
  \includegraphics[width=\textwidth]%
    {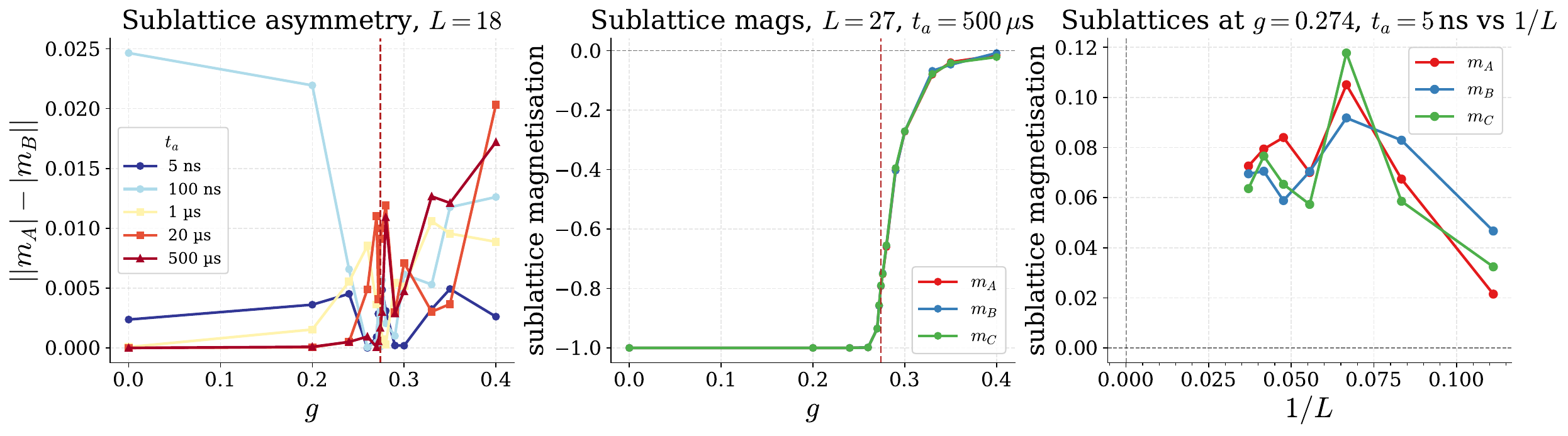}\\[2pt]
  \includegraphics[width=\textwidth]%
    {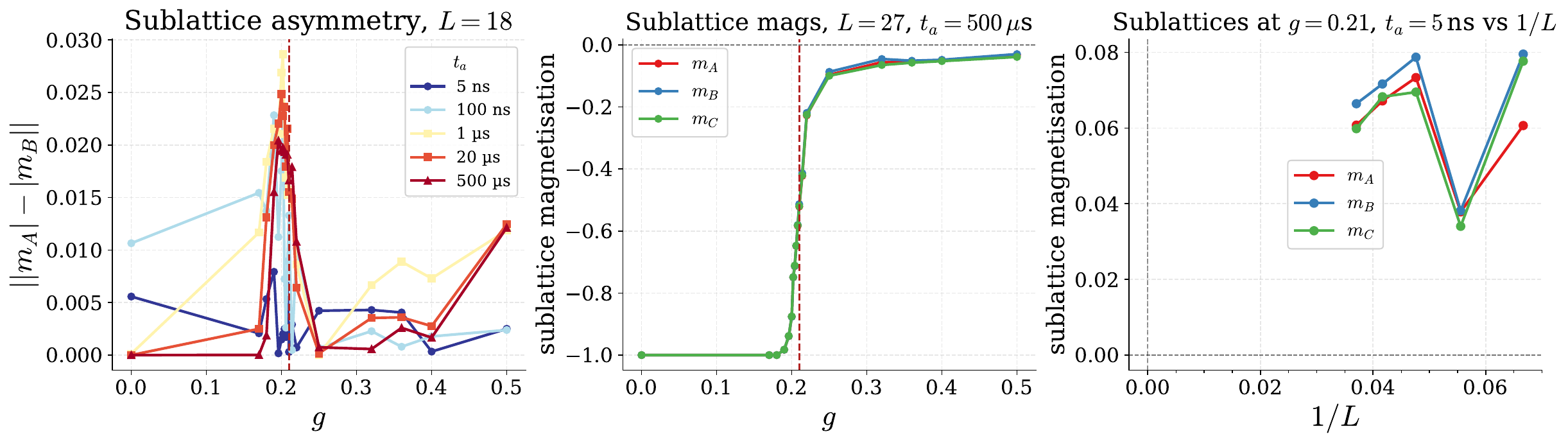}\\[2pt]
  \includegraphics[width=\textwidth]%
    {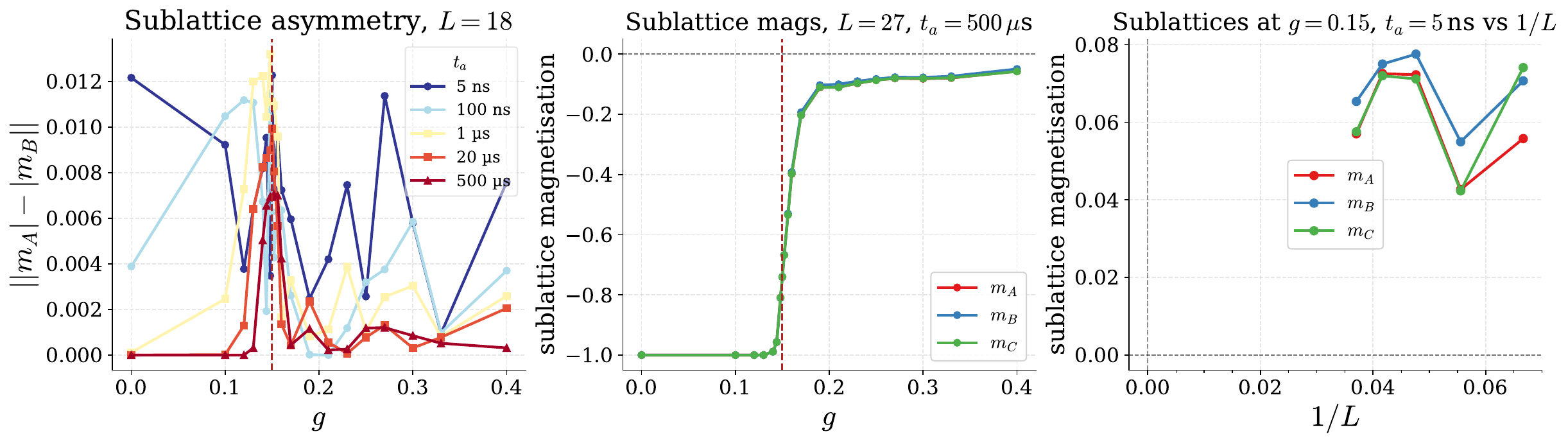}\\[2pt]
  \includegraphics[width=\textwidth]%
    {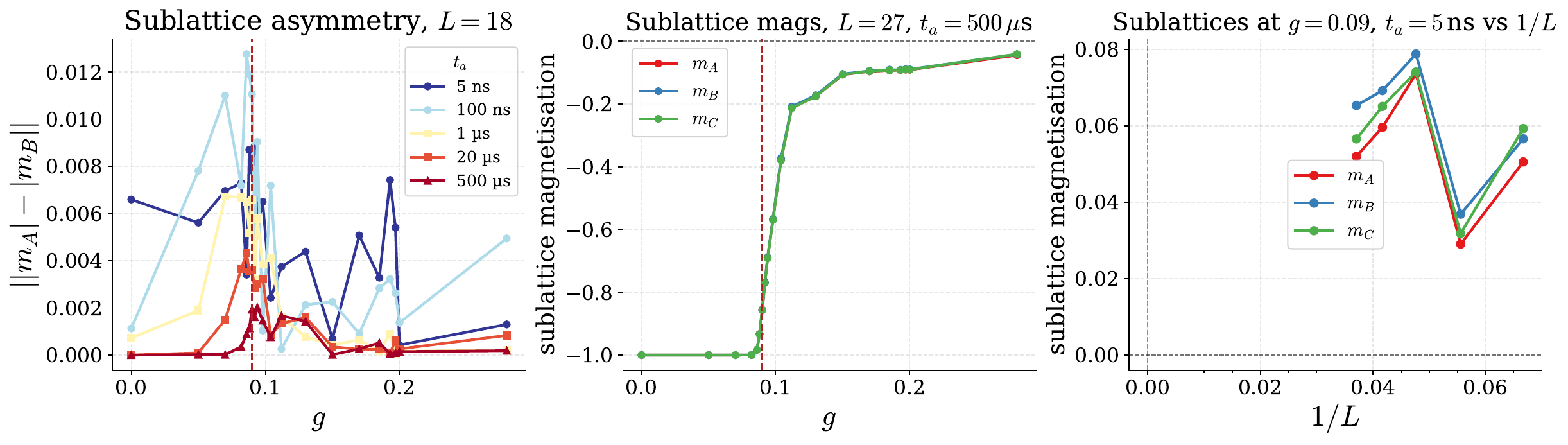}
  \caption{%
    \textbf{Direct FM-to-paramagnet transition across all four
    geometries.}
    Each row: (left) sublattice asymmetry $||m_A|-|m_B||$ vs $g$
    at $L=18$, confirming three-sublattice symmetry throughout;
    (centre) sublattice magnetisations $m_A$, $m_B$, $m_C$ vs $g$
    at $L=27$, $\ta=500\,\mu\mathrm{s}$, all three vanishing
    simultaneously at $\gcQPU(\alpha)$; (right) sublattice
    magnetisations vs $1/L$ at $g=\gcQPU(\alpha)$,
    $\ta=5\,\mathrm{ns}$, consistent with zero in the thermodynamic
    limit.
    No relative sublattice splitting at any $g$ or $\ta$ rules out
    an intermediate $\sqrt{3}\!\times\!\sqrt{3}$ phase.
    Top: $\alpha=1.0$. Second: $\alpha=0.7$. Third: $\alpha=0.5$.
    Bottom: $\alpha=0.3$.
  }
  \label{fig:sublattice}
\end{figure*}

For all four values of $\alpha$, $m_A$, $m_B$, and $m_C$ drop
simultaneously to zero at $\gcQPU(\alpha)$ with no relative
sublattice splitting at any $g$ or $\ta$ (Fig.~\ref{fig:sublattice}).
The $\sqrt{3}\!\times\!\sqrt{3}$ structure factor $S_{\sqrt{3}}/L^2$
shows no $L$-growth above $\gcQPU$, ruling out any intermediate
ordered phase.
Energy histograms at $\ta=500\,\mu\mathrm{s}$ are unimodal at all
$g$ near $\gcQPU$ across $L\in\{18,21,24,27\}$ for all four
geometries, consistent with continuous transitions.

The mechanism is $\alpha$-independent: on the triangular lattice,
each NNN bond shares two common NN bonds per
site~\cite{Villain1980,Moessner2001}, so partial AFM NNN ordering
necessarily disrupts FM NN exchange simultaneously, incurring a
large energy penalty.
This geometric constraint, rather than any specific coupling ratio,
enforces the direct transition for all $\alpha\in[0.3,1]$.
The theoretical phase diagram of Lee et al.~\cite{LeeKaulBalents2010}
argued for direct transitions in the FM-dominated sector on
perturbative grounds; the present QPU measurements confirm this
numerically at system sizes far beyond the reach of those methods.
The field-induced $\sqrt{3}\!\times\!\sqrt{3}$-type spin-flip phases
observed in CoNb$_2$O$_6$~\cite{Cabrera2014} require an applied field
for stabilisation and are absent at zero field, fully consistent
with the null $\sqrt{3}\!\times\!\sqrt{3}$ order found here.
The absence of incommensurate structure-factor growth confirms
the QPU measurements remain within the FM-dominated sector, well
below the IC phase boundary whose onset wavevector
$q_x^*=2\cos^{-1}(-\alpha/2)$ shifts with
$\alpha$~\cite{LeeKaulBalents2010}.

\subsection*{5.~Quantum ergodicity bypass and null VBS}

\begin{figure*}[t]
  \centering
  \includegraphics[width=0.95\textwidth]%
    {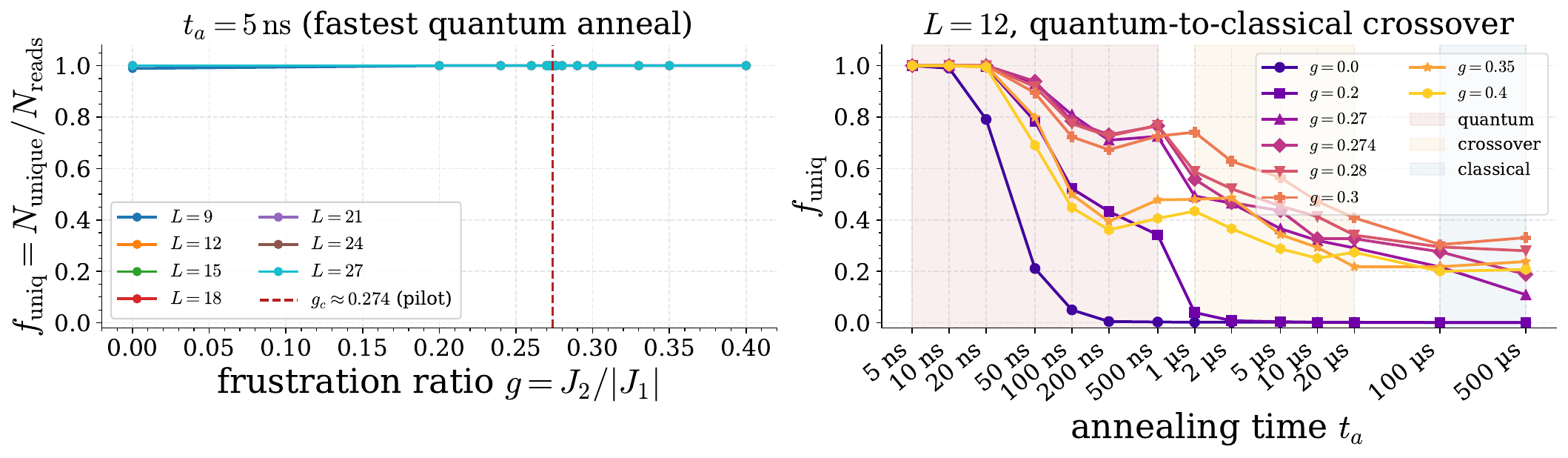}\\[2pt]
  \includegraphics[width=0.95\textwidth]%
    {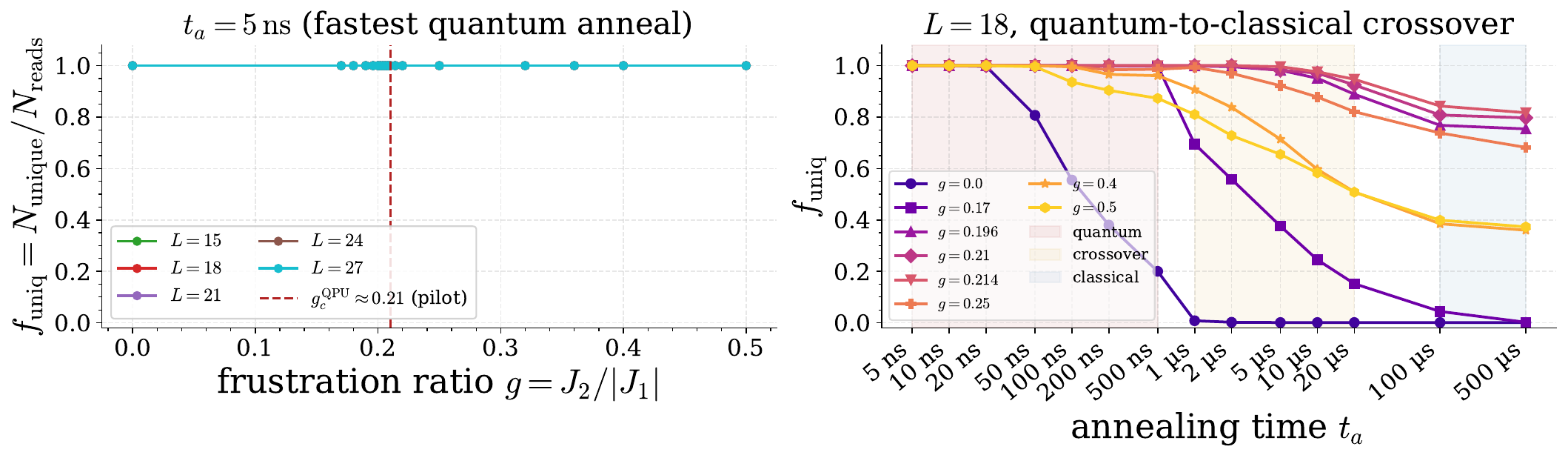}\\[2pt]
  \includegraphics[width=0.95\textwidth]%
    {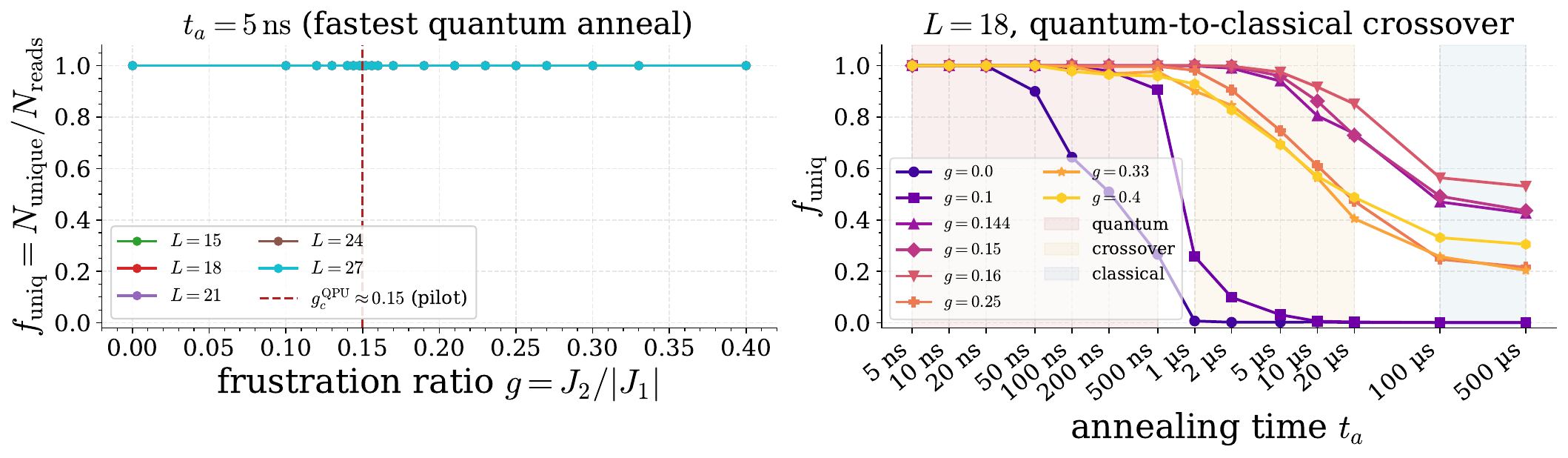}\\[2pt]
  \includegraphics[width=0.95\textwidth]%
    {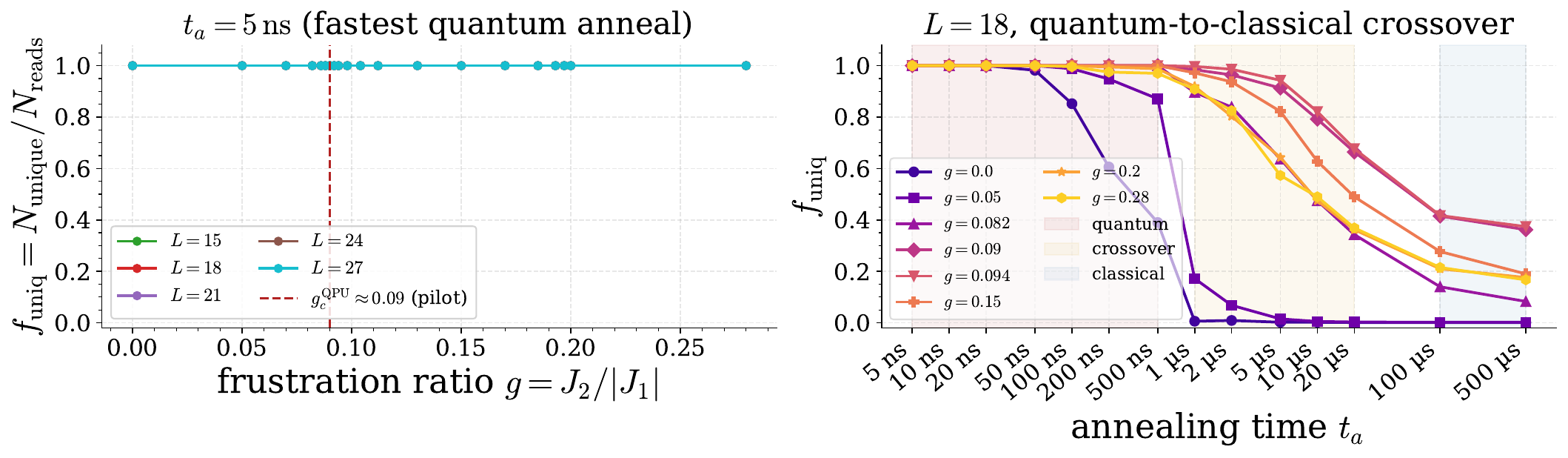}
  \caption{%
    \textbf{Quantum ergodicity bypass and quantum-to-classical
    crossover.}
    \textit{Left}: $f_{\rm uniq}=N_{\rm unique}/N_{\rm reads}$
    vs $g$ at $\ta=5\,\mathrm{ns}$.
    $f_{\rm uniq}=1.000$ for all $L$ and all $\alpha$: every anneal
    yields a distinct spin configuration, demonstrating complete
    ergodicity bypass via quantum tunnelling near each critical
    point.
    \textit{Right}: $f_{\rm uniq}$ vs $\ta$ for selected $g$
    ($L=12$ for $\alpha=1.0$; $L=18$ for $\alpha\in\{0.3,0.5,0.7\}$).
    The critical point $g\approx\gcQPU(\alpha)$ maintains
    $f_{\rm uniq}>0.5$ until $\ta\approx5\,\mu\mathrm{s}$, decaying
    more slowly than the FM anchor ($g=0$, frozen by
    $\ta\approx1\,\mu\mathrm{s}$), reflecting enhanced degeneracy
    at the quantum critical point.
    Top: $\alpha=1.0$. Second: $\alpha=0.7$. Third: $\alpha=0.5$.
    Bottom: $\alpha=0.3$.
  }
  \label{fig:funiq}
\end{figure*}

At $\ta=5\,\mathrm{ns}$, $f_{\rm uniq}=1.000$ for all $L$ and all
four $\alpha$ values (Fig.~\ref{fig:funiq}): every anneal produces
a distinct spin configuration, demonstrating quantum tunnelling
through the exponentially large configuration space near each
critical point.
These energy barriers have a direct experimental counterpart: Heid
et al.~\cite{Heid1995} report anomalously long relaxation times for
all field-induced transitions in CoNb$_2$O$_6$ at low temperatures,
attributed to the energy cost of reversing entire ferromagnetic
chains---the same barriers the QPU bypasses via quantum tunnelling
at $\ta=5\,\mathrm{ns}$.
At $g\approx\gcQPU(\alpha)$, $f_{\rm uniq}>0.5$ persists until
$\ta\approx5\,\mu\mathrm{s}$, decaying more slowly than the FM
anchor ($g=0$, frozen by $\ta\approx1\,\mu\mathrm{s}$), consistent
with enhanced degeneracy at each quantum critical point.
This behaviour is identical across all four geometries, confirming
ergodicity bypass is a property of the quantum fluctuations and the
frustrated critical point, not of any specific $\alpha$.
Plaquette order $\mathcal{O}_{\rm plaq}$ and string VBS order
$\mathcal{O}_{\rm VBS}$ extrapolate to zero as $L\to\infty$ at
all accessible $\ta$~\cite{Moessner2001}.

\section{Discussion}

\paragraph{Accessing a sign-problem model.}
The model in Eq.~\eqref{eq:H} cannot be simulated by quantum Monte
Carlo at any system size~\cite{Troyer2005}; exact diagonalisation
reaches at most $L=5$.
Prior QPU experiments on frustrated magnets operated exclusively in
the sign-problem-free antiferromagnetic sector, where classical
algorithms provide independent benchmarks~\cite{King2022,King2025,
Ali2024,Park2022}.
The present work occupies a fundamentally different regime: there is
no classical method to compare against, and the QPU results
constitute the first large-$L$ numerical data for this model family.
The plateau $\bar{r}=0.450\pm0.002$, confirmed independently across
three materials spanning a factor of more than two in coupling ratio,
is a genuinely new physical result---a universal ratio that no
perturbative or classical method could have predicted---complementing
the perturbative phase diagram of Lee et al.~\cite{LeeKaulBalents2010}
with the first quantitative phase boundaries at FSS-relevant system
sizes.

\paragraph{Measurement validity.}
Three mutually independent lines of evidence validate the QPU results
in the absence of a classical benchmark.
First, the zero chain-break fraction across all 5{,}572{,}000 shots
provides an empirical guarantee of embedding fidelity: every logical
spin is unambiguously defined on every anneal, with no residual
broadening of Binder crossings.
Second, the $\ta$-independence of $\gcQPU$ for
$\ta\gtrsim20\,\mu\mathrm{s}$ (Fig.~\ref{fig:phase}) confirms the
critical points are intrinsic to the model Hamiltonian.
Third, two successive sub-$1\sigma$ blind predictions and the
independent neutron-scattering corroboration of the CoNb$_2$O$_6$
phase boundary together establish the QPU as a quantitatively
reliable analogue computer for this class of sign-problem models.

\paragraph{Predictive power and experimental roadmap.}
Equation~\eqref{eq:prediction} provides the quantum phase boundary
for any member of the $M\mathrm{Nb_2O_6}$ and BaCo$_2$V$_2$O$_8$
families with a known structural anisotropy, without further QPU
measurements.
For CoNb$_2$O$_6$ ($\alpha\approx1$), inelastic neutron scattering
yields the dominant intrachain FM exchange $J_z=2.19\,\mathrm{meV}$
from spin-wave parametrisation of the full three-dimensional
dispersion~\cite{Cabrera2014} (a lower bound due to quantum
renormalisation of the dispersion bandwidth).
Our measurement $\gcQPU(1.0)=0.286\pm0.012$ then predicts the NNN
frustration at the quantum critical point as
$J_{2c}=0.286\,J_z\approx0.63\,\mathrm{meV}$, independently
corroborated by $J_2/J_1=0.76\pm0.10$ from the full spin-flip
dispersion~\cite{Cabrera2014}, which substantially exceeds
$\gcQPU(1.0)=0.286$ and confirms CoNb$_2$O$_6$ sits firmly in the
quantum-disordered sector.
For FeNb$_2$O$_6$ ($\alpha\approx0.3$), the reduced interchain
coupling places the material frustration ratio potentially close to
$\gcQPU(0.3)=0.093$, making it the columbite family member closest
to the quantum phase boundary and the most promising target for
polarised neutron diffraction seeking simultaneous vanishing of all
three sublattice magnetisations with no onset of
$\sqrt{3}\!\times\!\sqrt{3}$ order.
For BaCo$_2$V$_2$O$_8$, the predicted critical field in the
frustrated FM sector is distinct from the field-induced
incommensurate ordering~\cite{Kimura2008} and the one-dimensional
critical field~\cite{Faure2018}, which both occur in the
antiferromagnetic intrachain sector under a longitudinal field;
the present transition operates in a coupling regime orthogonal
to those studies.

\paragraph{Broader context.}
The strategy of using a QPU with minor embedding to extract
quantitative critical properties from a provably QMC-intractable
model is not specific to the frustrated Ising geometry.
A related demonstration for the Sherrington-Kirkpatrick spin glass
at up to $N=4000$ spins is given in Ref.~\cite{Ghosh2025RSB}.
Whether analogous approaches extend to models with itinerant degrees
of freedom depends on future advances in hardware connectivity;
the present study establishes the strategy's feasibility in the
spin-model sector.

\paragraph{Finite-size corrections and open questions.}
Non-monotonic Binder drift is most pronounced at $\alpha=1.0$
(pairwise spread $\delta\gc=0.025$, weighted uncertainty $\pm0.012$)
and smallest at $\alpha=0.7$ (inner-pair spread $\delta\gc=0.003$),
reflecting the stronger finite-size effects of the isotropic 2D
geometry.
The present $L\leq27$ exhausts the qubit budget of the Advantage2
processor for this embedding.
Three open questions require $L\geq45$: (i)~the universality class
($\nu\approx0.40$--$0.51$ across all four geometries, consistently
below $\nu_{\rm 3D\,Ising}=0.630$, suggesting a non-standard class
but inconclusive at current sizes); (ii)~the order of the
transitions (energy histograms are unimodal at all present sizes,
consistent with continuous transitions, but the Lee-Kosterlitz
diagnostic is inconclusive); and (iii)~the precise crossover scale
$\alpha^*$ and step amplitude $\Delta r$.

\section{Conclusion}

We have presented the first systematic large-scale numerical study
of the frustrated FM-AFM transverse-field $J_1$-$J_2$ Ising model
on the isosceles triangular lattice, using a D-Wave Advantage2 QPU
at $L\leq27$ across $\alpha\in\{0.3,0.5,0.7,1.0\}$.
Because the model has a provable sign problem, these are also the
first phase boundaries for this family at any FSS-relevant system
size.

The central result is a two-regime dimensional crossover in the
quantum suppression ratio.
Three independent quasi-1D geometries (FeNb$_2$O$_6$,
BaCo$_2$V$_2$O$_8$, NiNb$_2$O$_6$) are mutually consistent with a
universal plateau $\bar{r}=0.450\pm0.002$
($\chi^2/\mathrm{dof}=1.10$), demonstrating that quantum
fluctuations destroy approximately 55\% of the classical FM
stability window throughout the quasi-1D sector irrespective of
coupling anisotropy, with the plateau anchored by the 1D Pfeuty
fixed point ($r_0=0.494\pm0.024$, consistent with $r^{\rm 1D}=1/2$
within $1.7\sigma$).
At $\alpha>\alpha^*\approx0.7$ the system crosses over to 2D
behaviour; inner Binder cumulant pairs resolve a step
$\Delta r=0.038\pm0.015$ ($2.5\sigma$).
The linear fit $r(\alpha)=0.494-0.063\,\alpha$ ($1.9\sigma$)
summarises both regimes; its slope is a lower bound on the true
crossover amplitude, which is concentrated in $\alpha\in[\alpha^*,1]$,
is inherently nonlinear, and is further suppressed by the residual
isosceles distortion ($t_2/t_1=0.82$~\cite{Cabrera2014}) of the
physical $\alpha=1$ geometry.
Two sequential blind predictions confirmed at $0.2\sigma$ and
$0.7\sigma$, together with independent neutron-scattering validation
for CoNb$_2$O$_6$, establish the QPU as a reliable analogue
computer for this class of sign-problem models.

Equation~\eqref{eq:prediction} now provides a predicted quantum
phase boundary for any member of the $M\mathrm{Nb_2O_6}$ or
BaCo$_2$V$_2$O$_8$ families with a known structural anisotropy,
supplying concrete quantitative targets for neutron scattering
experiments.
Open questions---universality class, transition order, and the
precise crossover scale $\alpha^*$---await $L\geq45$ hardware.

\bibliographystyle{apsrev4-2}
\bibliography{ref_triangular}

@article{Cai2014,
  author  = {Cai, J. and Macready, W. G. and Roy, A.},
  title   = {A practical heuristic for finding graph minors},
  journal = {arXiv preprint arXiv:1406.2741},
  year    = {2014},
  url     = {https://arxiv.org/abs/1406.2741}
}

@article{Balents2010,
  author    = {Balents, Leon},
  title     = {Spin liquids in frustrated magnets},
  journal   = {Nature},
  volume    = {464},
  pages     = {199--208},
  year      = {2010},
  doi       = {10.1038/nature08917}
}

@article{Savary2017,
  author    = {Savary, Lucile and Balents, Leon},
  title     = {Quantum spin liquids: a review},
  journal   = {Rep. Prog. Phys.},
  volume    = {80},
  pages     = {016502},
  year      = {2017},
  doi       = {10.1088/0034-4885/80/1/016502} 
}

@book{Diep2005,
  editor    = {Diep, H. T.},
  title     = {Frustrated Spin Systems},
  publisher = {World Scientific},
  address   = {Singapore},
  year      = {2005},
  doi       = {10.1142/5697}
}

@book{Sachdev2011,
  author    = {Sachdev, Subir},
  title     = {Quantum Phase Transitions},
  edition   = {2nd},
  publisher = {Cambridge University Press},
  address   = {Cambridge},
  year      = {2011},
  doi       = {10.1017/CBO9780511973765}
}

@article{Troyer2005,
  author    = {Troyer, Matthias and Wiese, Uwe-Jens},
  title     = {Computational Complexity and Fundamental Limitations
               to Fermionic Quantum {Monte Carlo} Simulations},
  journal   = {Phys. Rev. Lett.},
  volume    = {94},
  pages     = {170201},
  year      = {2005},
  doi       = {10.1103/PhysRevLett.94.170201}
}

@article{Cabrera2014,
  author    = {Cabrera, I. and Thompson, J. D. and Coldea, R.
               and Prabhakaran, D. and Bewley, R. I.
               and Guidi, T. and Rodriguez-Rivera, J. A.
               and Stock, C.},
  title     = {Excitations in the quantum paramagnetic phase of the
               quasi-one-dimensional {Ising} magnet
               {CoNb$_2$O$_6$} in a transverse field: Geometric
               frustration and quantum renormalization effects},
  journal   = {Phys. Rev. B},
  volume    = {90},
  pages     = {014418},
  year      = {2014},
  doi       = {10.1103/PhysRevB.90.014418}
}

@article{LeeKaulBalents2010,
  author    = {Lee, Seung-Hun and Kaul, Ribhu K. and Balents, Leon},
  title     = {Interplay of quantum criticality and geometric
               frustration in columbite},
  journal   = {Nat. Phys.},
  volume    = {6},
  pages     = {702--706},
  year      = {2010},
  doi       = {10.1038/nphys1696}
}

@article{Coldea2010,
  author    = {Coldea, R. and Tennant, D. A. and Wheeler, E. M.
               and Wawrzynska, E. and Prabhakaran, D.
               and Telling, M. and Habicht, K.
               and Smeibidl, P. and Kiefer, K.},
  title     = {Quantum Criticality in an {Ising} Chain:
               Experimental Evidence for Emergent {E8} Symmetry},
  journal   = {Science},
  volume    = {327},
  pages     = {177--180},
  year      = {2010},
  doi       = {10.1126/science.1180085}
}

@article{Heid1995,
title = {Magnetic phase diagram of CoNb2O6: A neutron diffraction study},
journal = {Journal of Magnetism and Magnetic Materials},
volume = {151},
number = {1},
pages = {123-131},
year = {1995},
issn = {0304-8853},
doi = {https://doi.org/10.1016/0304-8853(95)00394-0},
url = {https://www.sciencedirect.com/science/article/pii/0304885395003940},
author = {C. Heid and H. Weitzel and P. Burlet and M. Bonnet and W. Gonschorek and T. Vogt and J. Norwig and H. Fuess}
}

@article{Kimura2008,
  author    = {Kimura, S. and Matsuda, M. and Masuda, T.
               and Hondo, S. and Kaneko, K. and Metoki, N.
               and Hagiwara, M. and Takeuchi, T. and Okunishi, K.
               and He, Z. and Yi, K. C. and Yoshida, Y.
               and Watanabe, S. and Kindo, K.},
  title     = {Collapse of Magnetic Order of the Quasi
               One-Dimensional {Ising}-Like Antiferromagnet
               {BaCo$_2$V$_2$O$_8$} in Transverse Fields},
  journal   = {Phys. Rev. Lett.},
  volume    = {101},
  pages     = {207201},
  year      = {2008},
  doi       = {10.1103/PhysRevLett.101.207201}
}

@article{Faure2018,
  author    = {Faure, Q. and Takayoshi, S. and Petit, S.
               and Simonet, V. and Raymond, S. and Regnault, L.-P.
               and Boehm, M. and White, J. S. and M{\aa}nsson, M.
               and R{\"u}egg, Ch. and Lejay, P. and Canals, B.
               and Lorenz, T. and Furuya, S. C.
               and Giamarchi, T. and Grenier, B.},
  title     = {Topological quantum phase transition in the
               {Ising}-like antiferromagnetic spin chain
               {BaCo$_2$V$_2$O$_8$}},
  journal   = {Nat. Phys.},
  volume    = {14},
  pages     = {716--722},
  year      = {2018},
  doi       = {10.1038/s41567-018-0126-8}
}

@article{Wannier1950,
  author    = {Wannier, Gregory H.},
  title     = {Antiferromagnetism. {The} Triangular {Ising} Net},
  journal   = {Phys. Rev.},
  volume    = {79},
  pages     = {357--364},
  year      = {1950},
  doi       = {10.1103/PhysRev.79.357}
}

@article{Pfeuty1970,
  author    = {Pfeuty, Pierre},
  title     = {The one-dimensional {Ising} model with a transverse
               field},
  journal   = {Ann. Phys.},
  volume    = {57},
  pages     = {79--90},
  year      = {1970},
  doi       = {10.1016/0003-4916(70)90270-8}
}

@article{Villain1980,
  author    = {Villain, J. and Bidaux, R. and Carton, J.-P.
               and Conte, R.},
  title     = {Order as an effect of disorder},
  journal   = {J. Phys. (Paris)},
  volume    = {41},
  pages     = {1263--1272},
  year      = {1980},
  doi       = {10.1051/jphys:0198000410110126300}
}

@article{Moessner2001,
  author    = {Moessner, R. and Sondhi, S. L. and Chandra, P.},
  title     = {Two-dimensional periodic frustrated {Ising} models
               in a transverse field},
  journal   = {Phys. Rev. B},
  volume    = {64},
  pages     = {144416},
  year      = {2001},
  doi       = {10.1103/PhysRevB.64.144416}
}

@article{King2022,
  author    = {King, Andrew D. and Suzuki, Sei and Raymond, Jack
               and Zucca, Alex and Lanting, Trevor
               and Altomare, Fabio and Berkley, Andrew J.
               and Eichhorn, Sara and Hoskinson, Emile
               and Johnson, Mark W. and others},
  title     = {Coherent quantum annealing in a programmable
               2{,}000 qubit {Ising} chain},
  journal   = {Nat. Phys.},
  volume    = {18},
  pages     = {1324--1328},
  year      = {2022},
  doi       = {10.1038/s41567-022-01741-6}
}

@article{King2025,
  author    = {King, Andrew D. and others},
  title     = {Beyond-classical computation in quantum simulation},
  journal   = {Science},
  volume    = {388},
  pages     = {199--204},
  year      = {2025},
  doi       = {10.1126/science.ado6285}
}

@article{Ali2024,
  author    = {Ali, A. and others},
  title     = {Quantum quench dynamics of geometrically frustrated
               {Ising} models},
  journal   = {Nat. Commun.},
  volume    = {15},
  pages     = {10756},
  year      = {2024},
  doi       = {10.1038/s41467-024-54701-4}
}

@article{park2022,
author = {Park ,Hayun and Lee ,Hunpyo},
title = {Frustrated Ising Model on D-wave Quantum Annealing Machine},
journal = {Journal of the Physical Society of Japan},
volume = {91},
number = {7},
pages = {074001},
year = {2022},
doi = {10.7566/JPSJ.91.074001}
}

@article{Albash2018,
  author    = {Albash, Tameem and Lidar, Daniel A.},
  title     = {Demonstration of a Scaling Advantage for a Quantum
               Annealer over Simulated Annealing},
  journal   = {Phys. Rev. X},
  volume    = {8},
  pages     = {031016},
  year      = {2018},
  doi       = {10.1103/PhysRevX.8.031016}
}

@article{Ghosh2025RSB,
      title={Exploring Replica Symmetry Breaking and Topological Collapse in Spin Glasses with Quantum Annealing}, 
      author={Kumar Ghosh},
      journal = {arXiv preprint arXiv:2511.06403 [cond-mat.dis-nn]},
      year={2025},
      url={https://arxiv.org/abs/2511.06403}, 
}

\end{document}